\documentclass[aps,pra,twocolumn,superscriptaddress,showkeys,amsmath,amssymb,longbibliography]{revtex4-1}
\usepackage{graphicx}
\usepackage{color}

\usepackage[colorlinks,urlcolor=blue,citecolor=blue,linkcolor=blue]{hyperref}
\usepackage[font=small,labelfont=bf,format=plain,justification=centerlast,labelsep=period]{caption}

\usepackage{orcidlink}

\begin{document} 

\title{Strain-spectroscopy of strongly interacting defects in  superconducting qubits} 

\author{Octavio de los Santos-S\'anchez\,\orcidlink{0000-0002-4316-0114}} 
\address{Escuela de Ingenier\'{\i}a y Ciencias, Instituto Tecnol\'ogico y de Estudios Superiores de Monterrey, \\  Avenida San Carlos 100, Campus Santa Fe, Ciudad de M\'exico, 01389, Mexico}

\author{Ricardo Rom\'an-Ancheyta\,\orcidlink{0000-0001-6718-8587}} 
\email{ancheyta6@gmail.com}
\address{Instituto Nacional de Astrof\'isica, \'Optica y Electr\'onica, Calle Luis Enrique Erro 1, Sta. Ma.  Tonantzintla, Puebla CP 72840, Mexico}

\date{\today}


\begin{abstract}
The proper functioning of some micro-fabricated novel quantum devices, such as superconducting resonators and qubits, is severely affected by the presence of parasitic structural material defects known as tunneling two-level-systems (TLS). Recent experiments have reported unambiguous evidence of the strong interaction between individual (coherent) TLS using strain-assisted spectroscopy. This work provides an alternative and simple theoretical insight that illustrates how to obtain the spectral response of such strongly interacting defects residing inside the amorphous tunnel barrier of a qubit's Josephson junction. Moreover, the corresponding spectral signatures obtained here may serve to quickly and efficiently elucidate the actual state of these interacting TLS in experiments based on strain- or electric-field spectroscopy. 
\end{abstract}


\maketitle


\section{Introduction} \label{sec:intro}

Impressive progress in the field of circuit quantum electrodynamics has occurred since the first realization of the coherent interaction between a superconducting qubit with a single microwave photon~\cite{Wallraff2004}. To date, such a coupled system constitutes the building block for the state-of-the-art prototypes of solid-state-based quantum processors~\cite{Google_2019}. Unfortunately, it is still challenging to preserve the qubits' coherence for a long time due to the unavoidable interaction with their environment~\cite{blais2020circuit,somoroff2021millisecond}, including cosmic rays~\cite{Martinis2021,Cardani2020}. A dominant source of decoherence is the random formation of natural defects at circuit interfaces and inside the Josephson junctions (JJs), whereby a qubit is highly sensitive~\cite{Murray2021,Martinis_PRL_2004,Martinis_PRL_2005,Burin_2014}. 

Superconducting qubits must operate at low temperatures, a condition under which amorphous solids, such as the typical oxide tunnel barrier of a JJ, exhibit a universal behavior explained by the so-called standard tunneling model (STM)~\cite{Lubchenko_2007}. This phenomenological model describes atomic-scale defects in amorphous materials that have long been construed as tunneling two-level systems (TLS)~\cite{Leggett_2013,esquinazi2013tunneling}. The presence of TLS affects, severely, the frequency range of operation and coherence time of superconducting qubits ~\cite{Martinis_PRL_2004,Yu2004,Driessen_PRB_2009}, so that several strategies to improve the qubits' quality have been implemented~\cite{Osman_2021,In_situ_2021,Earnest_2018}; for instance, the use of JJs of small area~\cite{Martinis_PRL_2005}. Nevertheless, the sensitivity to TLS makes superconducting qubits fitting devices to investigate individual TLS~\cite{Grabovskij232} in virtually arbitrary materials~\cite{Bilmes2021} and, hopefully, to be able to reveal their true microscopic origin~\cite{Muller_2019}. 
\textcolor{black}{On the other hand, earlier proposals used TLS as logical qubits~\cite{Zagoskin_2006} or quantum memories, as experimentally demonstrated in~\cite{Neeley2008}}

The possibility of the occurrence of strong interactions between TLS, formerly surmised in ~\cite{Burin1998}, has already been confirmed via a novel experimental technique for high-resolution strain-tuning spectroscopy~\cite{Lisenfeld2015}. In this context, through bending a qubit chip sample, defect spectroscopy data showed the distinctive avoided level crossings embedded in a very peculiar S-shaped signal~\cite{Lisenfeld2015}. More recently, with the same technique, electric fields were used to find the precise location of individual TLS on the chip circuit~\cite{Lisenfeld2019,Bilmes_SR_2020}.
A laborious task in these types of experiments is to search for and identify authentic signatures of mutual TLS interactions from the intricate plots of their spectroscopic data~\cite{Bilmes2021,Bilmes_Thesis,Bilmes2021probing}. Due to such difficulties, the present work aims to provide a theoretical description of some spectroscopic measurements associated with interacting TLS. 

In this work, we use the method of small rotations~\cite{Klimov_Sanchez,Klimov_JMO} to get an effective Hamiltonian that accurately describes the energy level structure of two strongly interacting TLS residing in the tunnel barrier of a qubit's JJ. We also propose using the Eberly-W\'odkiewicz (EW) spectrum~\cite{Eberly1977} as an alternative way to obtain the spectral response of those interacting TLS. In contrast to the standard Wiener–Khintchine power spectrum, computing the EW spectrum is more straightforward and does not require knowledge of a stationary state~\cite{CRESSER_1983}. The EW spectrum has recently been used to describe ultrastrongly coupled quantum systems~\cite{Salado_Mej_a_2021}, intermittent resonance fluorescence~\cite{RicTime}, and signatures of non-thermal baths in quantum thermalization processes~\cite{roman2019spectral}. 
Here, we obtain explicit expressions for the typical spectra of two coherently interacting TLS in qubit chips where strain or electric fields are applied. Our results show that the EW spectrum reproduces, reasonably well, the behavior of previously reported strain-spectroscopic data, including the characteristic S-shaped signal.

The structure of the paper is as follows. In Sec.~\ref{sec:model}, we make a brief description of the STM and present the Hamiltonian of two interacting TLS in the tunnel barrier of a qubit’s JJ. In Sec.~\ref{Energy_Spectrum}, we implement a unitary transformation to obtain an effective Hamiltonian with simplified eigenvalues and eigenstates. In Sec.~\ref{EW_specturm}, we evaluate the EW spectrum and obtain the spectral response of the corresponding coupled TLS. Finally, Sec.~\ref{Conclusions} outlines our conclusions. 



\section{Model} \label{sec:model}

\begin{figure}[t!]
\includegraphics[width=5.0cm, height=3.5cm]{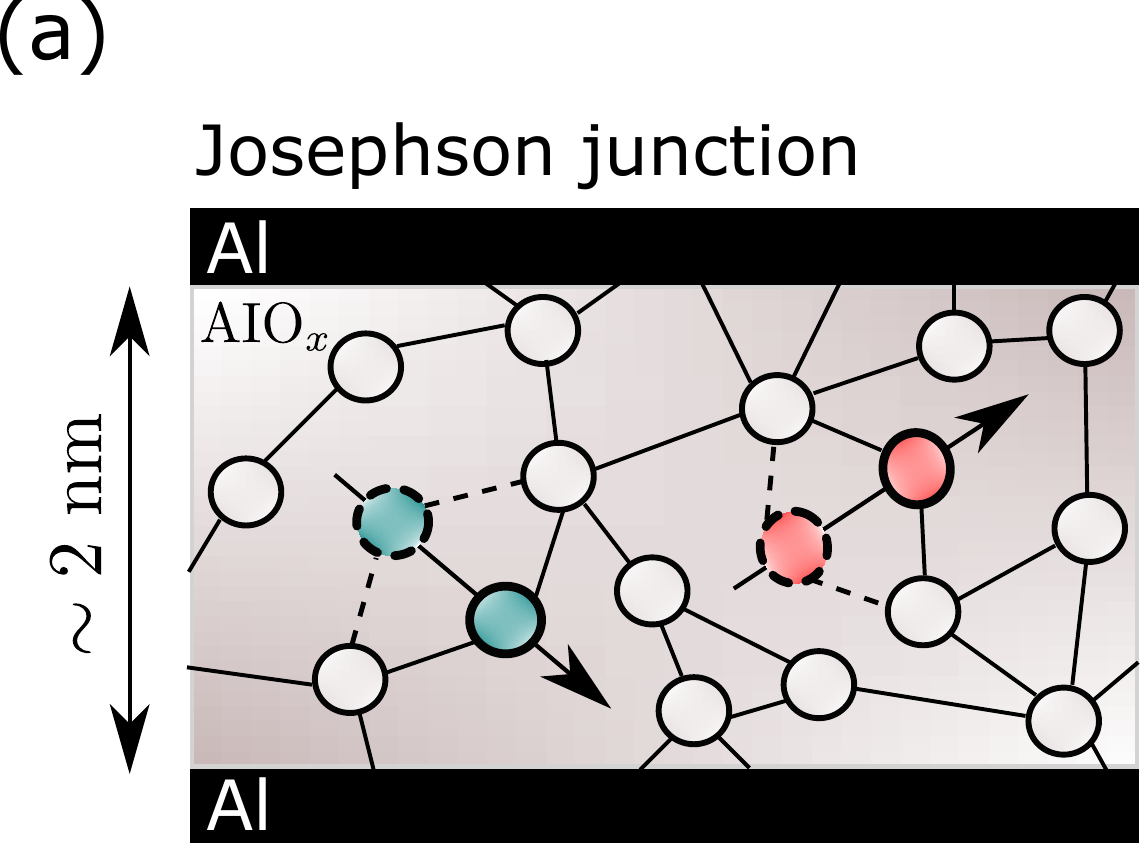}
\includegraphics[width=5.cm, height=4.cm]{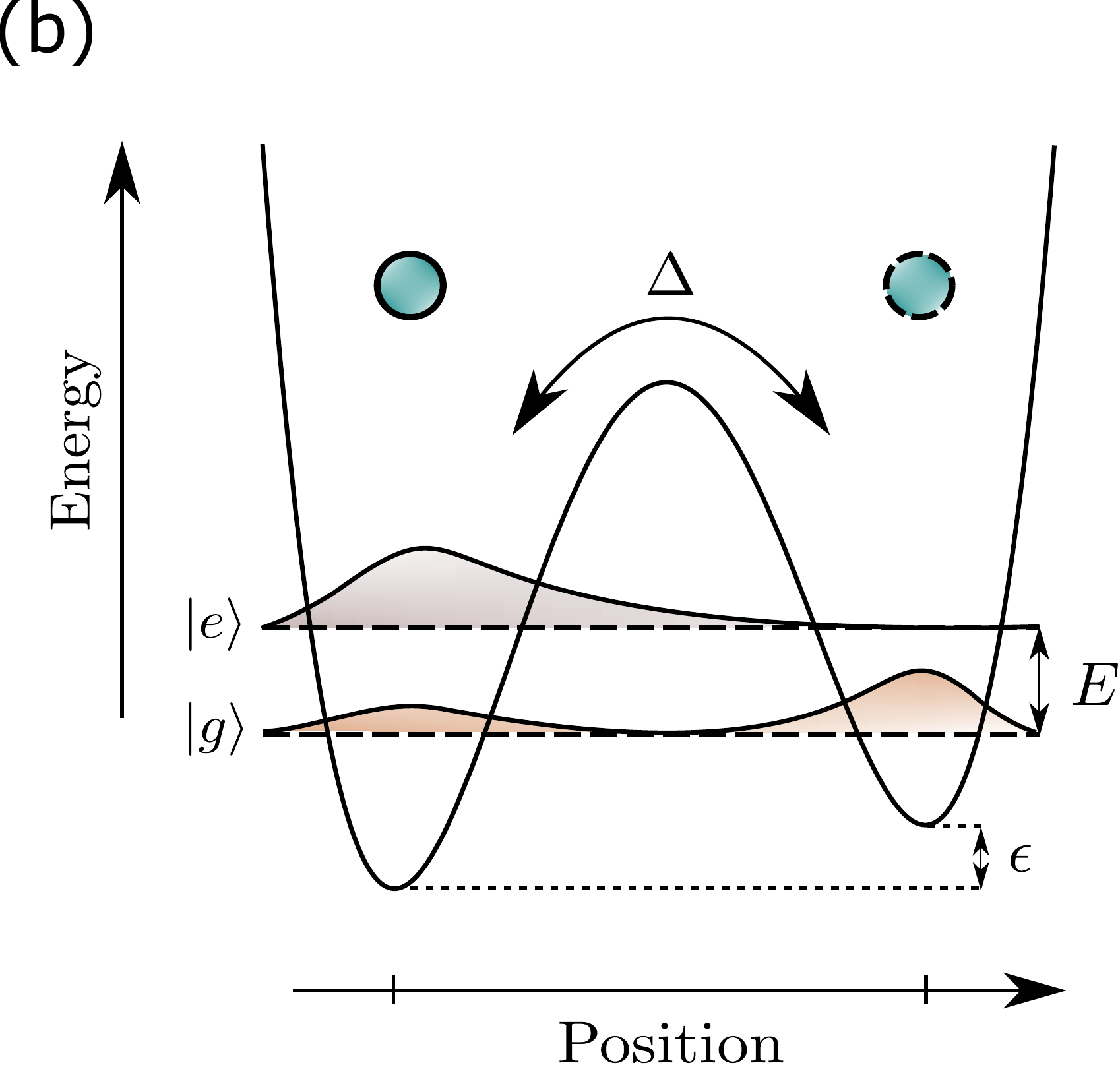} 
\caption{(a) Sketch of a Josephson tunnel junction consisting of two Al electrodes (black thin rectangles) isolated by a AlO$_{x}$ (amorphous aluminum oxide) dielectric layer with a typical thickness of $\sim$ 2 -- 3 nm. The latter hosts two defects viewed as atoms tunneling between two metastable locations (green and red circles); the arrows  denote their dipole moment. (b) Double-well potential model corresponding to a single atom tunneling between two metastable states. Tunneling and asymmetry energies, $\Delta$ and $\varepsilon$, respectively, set the transition energy $E =E_{e}-E_{g}= \sqrt{\Delta^{2}+\varepsilon^{2}}$ between the TLS eigenstates.}
\label{fig:potential}
\end{figure}

Our description is based upon the STM, whose visualization is that of an atom that can be sit in one of two parabolic potentials of a double-well, as depicted in Fig.~\ref{fig:potential}. Initially conceived by Phillips~\cite{Phillips1972} and Anderson~\cite{Halperin_1972} in the context of amorphous solids, the model is such that the minima of the wells represent energetically similar metastable states of an atom that are accessible via coherent tunneling. The inter-well tunneling energy $\Delta$ captures the capability of the atom to tunnel the potential barrier, and ${\varepsilon}$ is the energy asymmetry associated with the corresponding ground-state wavefunctions of the wells. Having specified the aforesaid atomic TLS parameters, which vary from defect to defect in amorphous solids \cite{Muller_2019}, the evolution of a single tunneling atom is ruled by the Hamiltonian (in units of $\hbar = 1$) \cite{Phillips1972,Halperin_1972}
\begin{equation}\label{eq:hamTLS}
H_\textsl{TLS} = \frac{1}{2}\varepsilon {\sigma}_{z}+\frac{1}{2}\Delta {\sigma}_{x},
\end{equation}
where ${\sigma}_{z} = |R\rangle \langle R|-|L\rangle \langle L|$ and ${\sigma}_{x} = |R\rangle \langle L|+|L\rangle \langle R|$ are the Pauli matrices written in the position or coordinate basis comprising of the left ($|L\rangle$) and right ($|R\rangle$) state vectors. The eigenstates of the Hamiltonian above, henceforth labeled as excited ($|e\rangle$) and ground ($|g\rangle$) states, are, respectively, $|e \rangle = \sin \left( \theta/2 \right) |L \rangle + \cos \left( \theta/2 \right) |R \rangle$, and $|g \rangle = \cos \left(\theta/2 \right) |L \rangle - \sin \left( \theta/2 \right) |R \rangle$, with $\tan \theta = \Delta / \varepsilon$. Within this diagonal basis, Hamiltonian (\ref{eq:hamTLS}) boils down to the simple form
\begin{equation}
H_{\textsl{TLS}} = \frac{1}{2} \omega \tilde{\sigma}_{z},
\end{equation}
where the tilde denotes Pauli operators in the eigenbasis and $\omega$ is the transition energy $E=E_e-E_g$ (also depicted in Fig.~\ref{fig:potential}) between eigenstates 
\begin{equation}
\omega =\sqrt{\varepsilon^{2}+\Delta^{2}}. \label{eq:splitting}
\end{equation}
Experimentally, the asymmetry energy $\varepsilon$ may be linearly driven by mechanical strain on the qubit chip, which, in turn, can be manipulated at will via an applied piezo voltage $V_{p}$ \cite{Muller_2019}; i.e., $\varepsilon = \varepsilon(V_{p})$ and, needless to say, the energy splitting will also be dependent on the strain field $\omega = \omega(V_{p})$. Strain tuning experiments in~\cite{Grabovskij232} confirm the hyperbolic dependence of (\ref{eq:splitting}) on the asymmetry energy $\varepsilon$, providing firm evidence for a central hypothesis that TLS are the cause of avoided level crossings in the frequency of JJ qubits.

Strain interactions between TLS occur by exchanging phonons~\cite{Carruzzo_2021,Klein_PRB_1978}.
In order to go into the scenario of two interacting TLS (labeled as TLS1 and TLS2) residing in the tunnel barrier of a JJ, the following Hamiltonian has been posited by Lisenfeld et al.~\cite{Lisenfeld2015} $H_{T} =  H_{1}+H_{2}+H_\texttt{TLS-TLS}$,
in which 
$2H_{i}=\varepsilon_{i}(V_{p}) {\sigma}_{z}^{(i)}+\Delta_{i} {\sigma}_{x}^{(i)}$
represents the Hamiltonian of a single defect $i$ and the TLS-TLS interaction, $H_\texttt{TLS-TLS}$, is taken to be described in the position basis as~\cite{Muller_2019}:
\begin{equation}
H_\texttt{TLS-TLS} = \frac{1}{2}g {\sigma}_{z}^{(1)}{\sigma}_{z}^{(2)},
\end{equation}
with $g$ being the strength of the defects' coupling whose value can be inferred from direct spectroscopic measurements. In this work, as in~\cite{Lisenfeld2015}, we find it convenient to reframe Hamiltonian $H_T$ by deploying the transformation ${\sigma}_{z}^{(i)} \rightarrow \cos\theta_{i}\,\tilde{\sigma}_{z}^{(i)}+\sin\theta_{i}\,\tilde{\sigma}_{x}^{(i)}$ and then neglecting coupling terms of the form $ \propto \tilde{\sigma}_{x}^{(1)}\tilde{\sigma}_{z}^{(2)}$ and $\propto \tilde{\sigma}_{z}^{(1)}\tilde{\sigma}_{x}^{(2)}$, because they contribute as small energy offsets. Thus, the following Hamiltonian is obtained in the diagonal basis
\begin{equation}
H_{T} = \frac{\omega_{1}}{2} \tilde{\sigma}_{z}^{(1)}+ \frac{\omega_{2}}{2} \tilde{\sigma}_{z}^{(2)}+\frac{g_{\parallel}}{\textcolor{black}{2}} \tilde{\sigma}_{z}^{(1)} \tilde{\sigma}_{z}^{(2)}+\frac{g_{\perp}}{2} \tilde{\sigma}_{x}^{(1)}\tilde{\sigma}_{x}^{(2)}.
\label{eq:ham1}
\end{equation}

Here, $\omega_{i}=\sqrt{\varepsilon_{i}^{2}(V_{p})+\Delta_{i}^{2}}$, $i$=$1,2$; $\textcolor{black}{g_{\parallel}}= g\cos \theta_{1} \cos \theta_{2}$ and $g_{\perp} = g \sin \theta_{1} \sin \theta_{2}$ are referred to as, respectively, the longitudinal and transversal coupling components, with $\cos \theta_{i}=\varepsilon_{i}/\omega_{i}$ and $\sin \theta_{i} = \Delta_{i}/\omega_{i}$. So, each TLS is characterized by the tunneling energy $\Delta_{i}$ and the strain-dependent asymmetry energy $\varepsilon_{i}(V_{p})$. The third and fourth terms appearing in~(\ref{eq:ham1}) are responsible for frequency shift and energy exchange effects, respectively, accounting for the spectral profile of the defect interaction already experimentally reported \cite{Lisenfeld2015}, as we will see below. Again, we remark that the asymmetry energy referred to above can be tuned by slightly bending the chip circuit using a piezo actuator, according to the rule $\varepsilon_{i} = c_{i} (V_{p}-V_{0i})$~\cite{Grabovskij232}; $V_{0i}$ is the voltage at which the TLS is at its symmetry point and $c_{i}$ is an adjusting parameter. 

\section{Energy spectrum}\label{Energy_Spectrum}

Hamiltonian~(\ref{eq:ham1}) has been successful in predicting, numerically, the transition energies of interacting TLS in a phase qubit~\cite{Lisenfeld2015}. Here we find that an algebraic diagonalization of it is possible, allowing us to arrive at the following closed-form expressions for its four eigenenergies:    
\begin{subequations}
\begin{eqnarray}
E_{1,2} & = & \textcolor{black}{ \mp\frac{1}{2}  \sqrt{(\omega_1+\omega_2)^{2}+{g}_{\perp}^{2}}+\frac{g_\parallel}{2} }, \label{eq:efull1}\\
E_{3,4} & = & \textcolor{black}{ \pm \frac{1}{2}  \sqrt{(\omega_1-\omega_2)^{2}+{g}_{\perp}^{2}} -\frac{g_\parallel}{2}. } \label{eq:efull4}
\label{eq:efullpm}
\end{eqnarray}
\end{subequations}
Associated with these eigenenergies, the corresponding eigenstates are given, in the basis $\{ |e,e\rangle, |e,g\rangle, |g,e\rangle, |g,g\rangle \}$,  respectively, by
\begin{subequations}
\begin{eqnarray}
|u_{1} \rangle & = &  \cos ({\alpha}/{2})|g,g \rangle-\sin ({\alpha}/{2}) |e,e \rangle, \\
|u_{2} \rangle & = &  \cos ({\alpha}/{2}) |e,e \rangle + \sin ({\alpha}/{2})|g,g \rangle, \\
|u_{3} \rangle & = &  \cos  ({\beta}/{2}) |e,g \rangle +  \sin  ({\beta}/{2})|g,e \rangle, \\
|u_{4} \rangle & = &  \cos ({\beta}/{2} )|g, e \rangle - \sin ({\beta}/{2}) |e,g \rangle,
\end{eqnarray}
\end{subequations}
where $\tan \alpha =\textcolor{black}{ {g}_{\perp}/(\omega_1+\omega_2)}$ and $\tan \beta =  \textcolor{black}{ {g}_{\perp}/(\omega_1-\omega_2)}$. Besides this result, one of the main contributions of this work is to remark the fact that, even in the strong coupling regime, Hamiltonian (\ref{eq:ham1}) can be further simplified by applying the transformation $T= e^{\eta {G}}$, with ${G} = \tilde{\sigma}_{+}^{(1)} \tilde{\sigma}_{+}^{(2)}-\tilde{\sigma}_{-}^{(1)} \tilde{\sigma}_{-}^{(2)}$ and choosing $\eta = g_{\perp}/[2(\omega_{1}+\omega_{2})]$ \cite{Klimov_Sanchez,Klimov_JMO}. Accordingly, keeping second-order terms involving $\eta$, we get the effective Hamiltonian (${H}_{\rm eff} = {T}{H_T}{T}^{\dagger}$):
\begin{eqnarray}
{H}_{\textrm{eff}} & \approx & \frac{\bar{\omega}_{1}}{2} \tilde{\sigma}_{z}^{(1)}+ \frac{\bar{\omega}_{2}}{2} \tilde{\sigma}_{z}^{(2)}+\frac{g_{\parallel}}{\textcolor{black}{2}} \tilde{\sigma}_{z}^{(1)} \tilde{\sigma}_{z}^{(2)} \nonumber \\
& & \qquad\qquad\qquad+\frac{g_{\perp}}{2} \big(\tilde{\sigma}_{+}^{(1)}\tilde{\sigma}_{-}^{(2)}+\tilde{\sigma}_{-}^{(1)}\tilde{\sigma}_{+}^{(2)}\big), \label{eq:heff}
\end{eqnarray}
where the energy exchange term, weighted by $g_{\perp}$, is now reminiscent of that of two dipole coupled two-level systems in the rotating-wave approximation, and 
\textcolor{black}{
we have defined 
\begin{equation}
\bar{\omega}_{j}=\omega_{j}+\frac{g_{\perp}^{2}}{4(\omega_{1}+\omega_{2})}.   
\end{equation} }
So, the four eigenvalues obtained from this effective Hamiltonian take the form
\begin{subequations}
\begin{eqnarray}
E^{\rm eff}_{1,2} & = & \textcolor{black}{+\frac{g_\parallel}{2} \mp \frac{1}{2}(\bar\omega_1+\bar\omega_2)}, \label{eq:eff1} \\
E^{\rm eff}_{3,4} & = & \textcolor{black}{-\frac{g_\parallel}{2} \pm\frac{1}{2} \sqrt{(\bar\omega_1-\bar\omega_2)^{2}+{g}_{\perp}^{2}} }. \label{eq:eff4}
\end{eqnarray}
\end{subequations}
Correspondingly, the eigenstates are also simplified as follows:
\begin{subequations}
\begin{eqnarray}
|{u}_{1} \rangle_{\rm eff} & = & |g,g\rangle,\\ 
|{u}_{2} \rangle_{\rm eff} & = & |e,e\rangle,  \\
|{u}_{3} \rangle_{\rm eff} & = & \cos ({\beta}/{2}) |e,g\rangle + \sin ({\beta}/{2})|g,e\rangle, \\
|{u}_{4} \rangle_{\rm eff} & = & \cos ({\beta}/{2}) |g,e\rangle - \sin ({\beta}/{2}) |e,g\rangle.
\end{eqnarray}
\end{subequations}
\begin{figure}[t]
\includegraphics[scale=0.4]{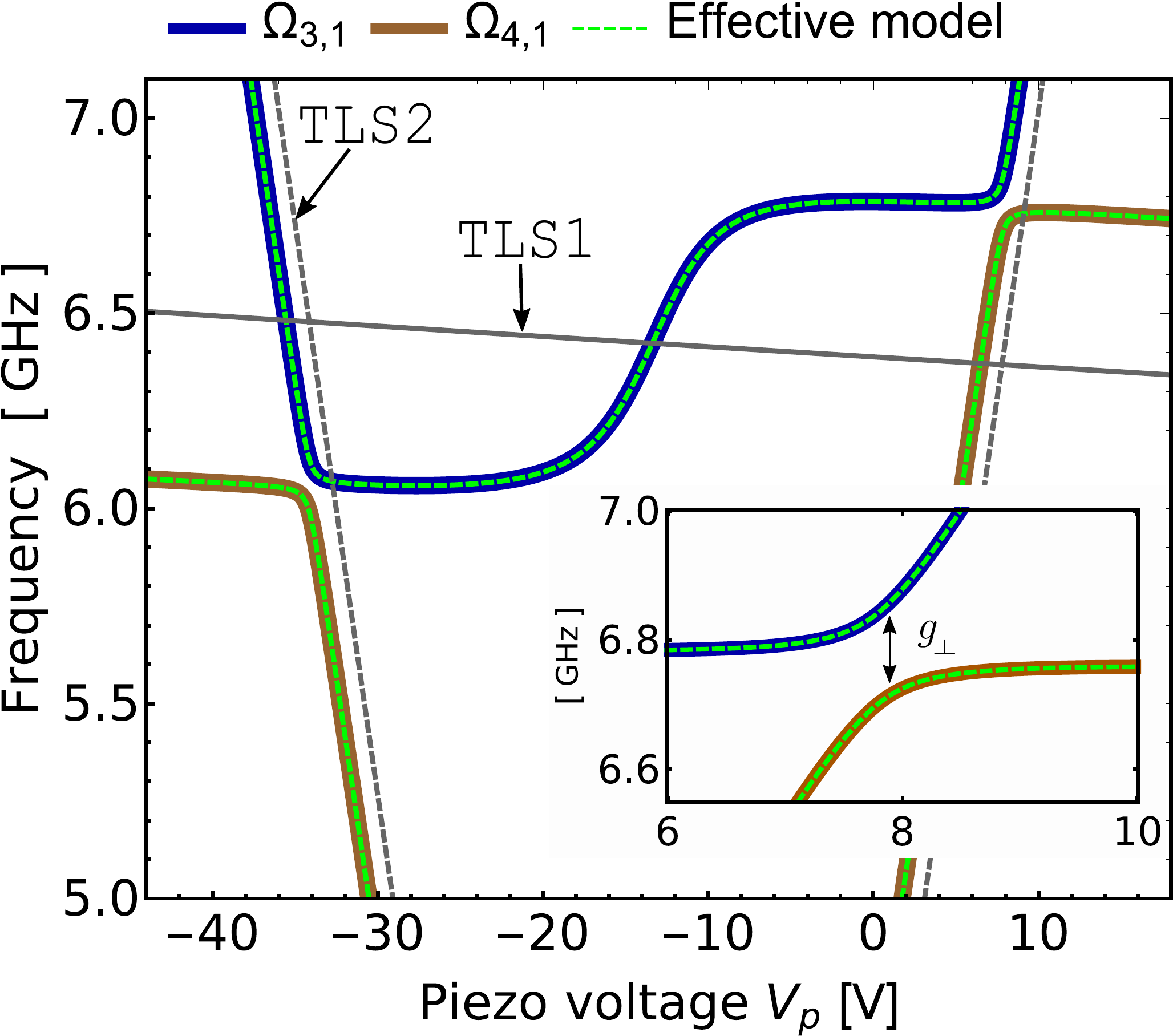}
\caption{Energy spectrum, $\Omega_{k,1} =E_{k}-E_{1}$, with $k=3,4$, obtained by diagonalizing the total (\ref{eq:ham1}) and effective (\ref{eq:heff}) Hamiltonian, vs. the piezo voltage, $V_{p}$. Gray lines represent the bare frequency of TLS1 (continuous) and TLS2 (dashed). System parameters: $g=800$ MHz, $\Delta_{1}=5.47$ GHz and $\Delta_{2}=1.3$ GHz; $\varepsilon_{1} = c_{1}V_{p}-3.3 [\textrm{GHz}]$ and $\varepsilon_{2}(V_{p}) = c_{2} (V_{p}+13 [\textrm{V}])$;  $c_{1}=5$ MHz/V and $c_{2} = 0.3$ GHz/V. The inset zooms in on the vicinity of the right avoided level crossing \textcolor{black}{where $V_{p} \sim 7.78$~V, $g_{\perp} \sim 140$ MHz and  $g_{\parallel} \sim -401$ MHz.}}
\label{fig:Elevels}
\end{figure}
In Fig.~\ref{fig:Elevels}, we display a comparison between the outcome of the total Hamiltonian~(\ref{eq:ham1}) (continuous lines) and that of the effective one~(\ref{eq:heff}) (green dashed lines) by plotting the strain dependence of the transition energies $ \Omega_{k,1} \equiv E_{k}-E_{1} $, for $k = 3,4$, making use of Eqs.~(\ref{eq:efull1})--(\ref{eq:efull4}) and (\ref{eq:eff1})--(\ref{eq:eff4}). 
\textcolor{black}{The transition energies may occur between the entangled states, $|{u}_{3} \rangle_{\rm eff}$ and $|{u}_{4}\rangle_{\rm eff}$, with the state $|u_1\rangle_{\rm eff}$.} This identification permits us to assess, for instance, the transversal coupling $g_{\perp}$ that takes part in the defect interplay:  near the anti-crossing region, $\omega_{1} \sim \omega_{2}$ (i.e., at resonant locus where the avoided level crossings take place) spectroscopic techniques enable one to trace the energy splitting $\Omega_{3,1}-\Omega_{4,1} =\textcolor{black}{ [{(\omega_1-\omega_2)^{2}+{g}_{\perp}^{2}}]^{1/2}} \sim g_{\perp}$. The system parameters used in the present calculation are taken to be close to the data formerly reported \cite{Lisenfeld2015}: the interplay coupling strength $|g| = 800$ MHz; as to tunneling energies, $\Delta_{1} = 5.47$ GHz, $\Delta_{2}=1.3$ GHz; the asymmetry energies are such that $\varepsilon_{1} = c_{1}V_{p}-3.3 [\textrm{GHz}]$ and $\varepsilon_{2}(V_{p}) = c_{2} (V_{p}+13 [\textrm{V}])$, with $c_{1}=5$ MHz/V and $c_{2} = 0.3$ GHz/V; the choice of these parameters is such that, in the displayed domain, TLS1 is barely affected by the mechanical strain. With the help of the foregoing information, we get $g_{\perp} \sim 0.14$ GHz, the two avoided level crossings are found to be positioned at $V_{p,\pm} \sim \{-34.15, \ 7.78 \} [V]$, and, say, at the right resonant location, the normalized coupling $g_{\perp}/\omega_{1,2}\sim 0.022$ is obtained; this makes legitimate the use of the effective Hamiltonian (\ref{eq:heff}) conserving solely the rotating terms~\cite{FriskKockum2019}. So, regarding the energy spectrum, we confirm that our effective model (green dashed lines) makes essentially the same prediction as the total  Hamiltonian (dark blue and brown solid lines). 
\textcolor{black}{In the next section,  $H_{\rm eff}$ will be helpful in obtaining the spectral response of the interacting TLSs.
}

\section{Strain-dependent spectra}\label{EW_specturm}

Customarily, one would describe the system's open dynamics under study by establishing a fitting local~\cite{Correa_Quantum_2021}, or global~\cite{Cattaneo_2019}, master equation. From this standpoint, knowledge of the steady-state of the system would be necessary to compute, for instance, the stationary power spectrum \`a la Wiener-Khintchine to attempt to trace its spectroscopy. However, as far as we know, information about the intrinsic damping mechanisms within amorphous materials involved in mutual TLS coherent interactions is neither obvious~\cite{Lisenfeld2016SR,ElectronicDecoherence} nor accessible, making it quite challenging to posit an accurate microscopic master equation treatment~\cite{Correa_OSIF_2017}. In order to sidestep this issue, we resort to carrying out a spectral measurement based upon the operational definition of spectrum for non-stationary processes, i.e., the so-called time-dependent ``physical spectrum'' of Eberly and W\'odkiewicz \cite{Eberly1977}. In fact, the EW spectrum has recently and successfully described the ultrastrong coupling between light and matter~\cite{Salado_Mej_a_2021} and even quantum thermalization processes~\cite{roman2019spectral}. Remarkably, the EW spectrum will also permit us to concisely illustrate the spectral response of strongly    interacting TLS in experiments based on strain-field spectroscopy. The standard definition of it is given by~\cite{Eberly1977,Eberly1980}
\begin{equation}\label{eq:ebspectrum}
\begin{split} 
S(\omega,\Gamma,t)  =  2\Gamma \int_{0}^{t}dt_{1}&\int_{0}^{t}dt_{2}  e^{-(\Gamma+i\omega)(t-t_{2})} \\
& \quad\times e^{-(\Gamma-i\omega)(t-t_{1})}G(t_{1},t_{2}),  
\end{split}
\end{equation}
where $\Gamma$ is the band half-width of a given interferometer acting as a filter and $\omega$ its central frequency. $G(t_{1},t_{2})$ is the corresponding two-time autocorrelation function that for the $i^{\rm th}$ TLS reads $G(t_{1},t_{2})= \langle \tilde{\sigma}_{+}^{(i)}(t_{1})\tilde{\sigma}_{-}^{(i)}(t_{2}) \rangle$, where the expectation value $\langle\cdots\rangle$ is taken with respect to some initial state of the two coupled TLS, and $\tilde{\sigma}_\pm^{(i)}(t_j)=\exp(iH_Tt_j)\tilde{\sigma}_\pm^{(i)}\exp(-iH_Tt_j)$.
Having diagonalized the total Hamiltonian~(\ref{eq:ham1}), one can obtain a closed-form expression for $\tilde{\sigma}_\pm(t_j)$ and, consequently, for $G(t_1,t_2)$.

\begin{figure}[t]
\includegraphics[scale=0.4]{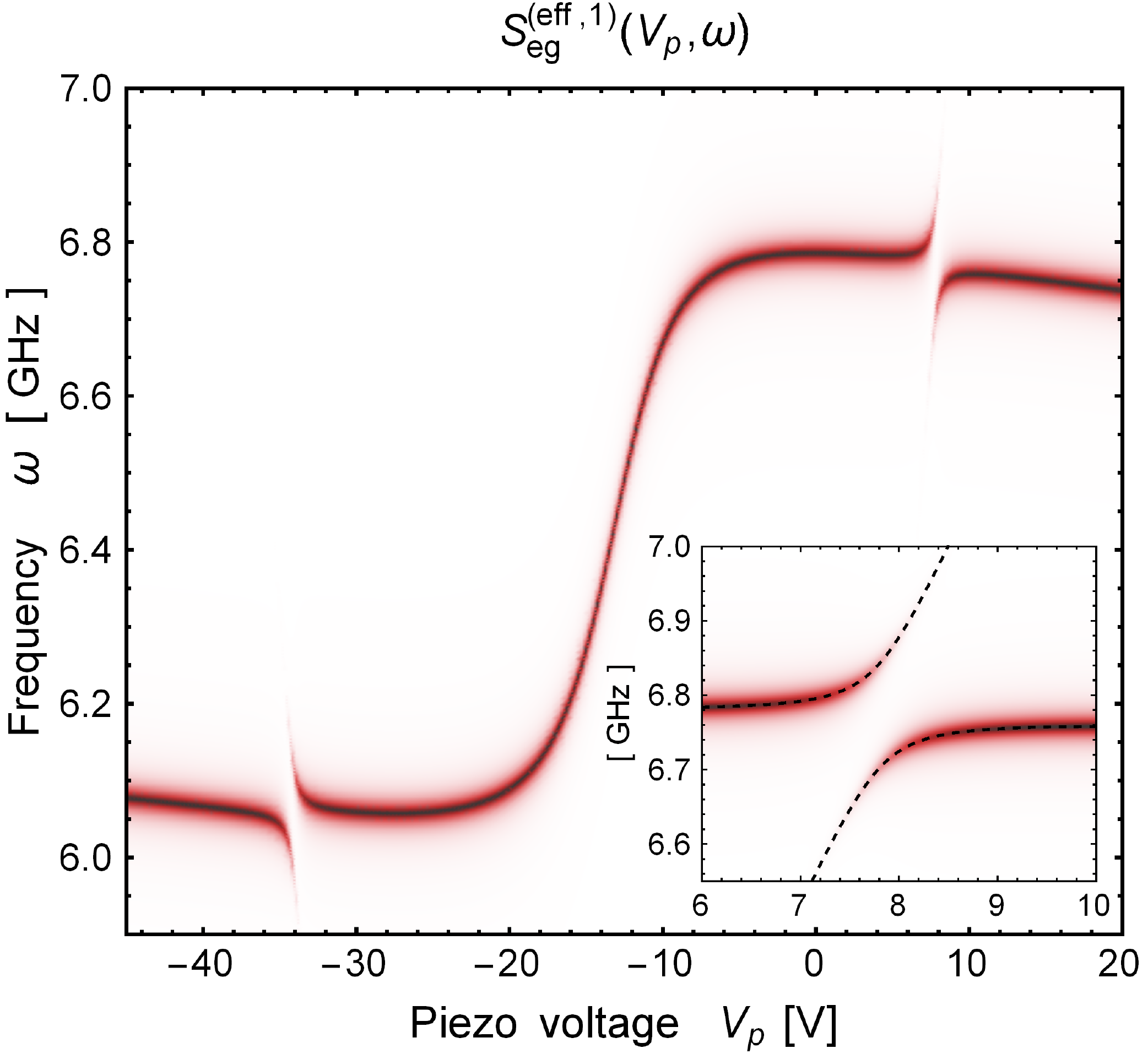}
\caption{Strain-dependent spectrum of TLS1 given by Eq.~(\ref{eq:effspectrum1}). The ordinate and abscissa indicate, respectively, the spectral detuning and piezo voltage $V_{p}$. The S-shaped profile and avoided level crossings are conspicuous. The inset zooms in on the same spectral signature in the neighbourhood of the right avoided level crossing at $\sim 7.8$ V. The initial state is taken to be the $|e,g\rangle$ state and the system parameters are the same as in Fig.~\ref{fig:Elevels} with $\Gamma=0.01$GHz.} 
\label{fig:scheme4}
\end{figure}

If we assume that the initial state of the composite system is $|e,g\rangle=|e\rangle\otimes|g\rangle$, the autocorrelation function takes the form
\begin{eqnarray} \label{eq:corr}
G_{eg}(t_{1},t_{2}) & = & \langle e,g| \tilde{\sigma}_{+}^{(1)}(t_{1})\tilde{\sigma}_{-}^{(1)}(t_{2}) |e,g\rangle, \nonumber \\
& = &  \cos^{2}(\alpha/2)\cos^{4}(\beta/2)e^{i \Omega_{3,1}(t_{1}-t_{2})}  \nonumber \\  
& & + \cos^{2}(\alpha/2)\sin^{4}(\beta/2)e^{i \Omega_{4,1}(t_{1}-t_{2})}  \nonumber \\
& & + \sin^{2}(\alpha/2)\cos^{4}(\beta/2)e^{i \Omega_{3,2}(t_{1}-t_{2})} \nonumber \\
& & + \sin^{2}(\alpha/2)\sin^{4}(\beta/2) e^{i\Omega_{4,2}(t_{1}-t_{2})} \nonumber \\
& & + \cos^{2}(\alpha/2) \sin^{2}(\beta) e^{i(\Omega_{3,1}t_{1}-\Omega_{4,1}t_{2})}/4 \nonumber \\
& & + \cos^{2}(\alpha/2) \sin^{2}(\beta) e^{i(\Omega_{4,1}t_{1}-\Omega_{3,1}t_{2})}/4 \nonumber \\
& & + \sin^{2}(\alpha/2)\sin^{2}(\beta) e^{i(\Omega_{3,2}t_{1}- \Omega_{4,2}t_{2})}/4 \nonumber \\
& & + \sin^{2}(\alpha/2)\sin^{2}(\beta) e^{i(\Omega_{4,2}t_{1}-\Omega_{3,2}t_{2})}/4.\qquad 
\end{eqnarray}
Here, the subscript ``$eg$'' indicates that TLS1 is initially excited and the TLS2 is in its ground state. Superscript ``$(1)$'' above the spin operators denotes that the autocorrelation and the EW spectrum are, in this case, associated with the TLS1. Notice that the time-dependent functions in (\ref{eq:corr}) are just simple exponentials; thus, their corresponding time integrals will be straightforward. So, on substitution of (\ref{eq:corr}) into (\ref{eq:ebspectrum}), once the transient process has elapsed, i.e., 
in the long time limit $\Gamma t \gg 1 $~\cite{RicTime}, this filtering integration yields the following expression for the TLS1's physical spectrum
\begin{eqnarray}
S_{eg}^{(1)}&&(\omega) = \nonumber \\ 
&& \frac{2\Gamma \cos^{2}(\alpha/2)\cos^{4}(\beta/2)}{\Gamma^{2}+(\omega-\Omega_{3,1})^{2}}  + \frac{2\Gamma \cos^{2}(\alpha/2)\sin^{4}(\beta/2)}{\Gamma^{2}+(\omega-\Omega_{4,1})^{2}} 
\nonumber\\ && + 
\frac{2\Gamma \sin^{2}(\alpha/2)\cos^{4}(\beta/2)}{\Gamma^{2}+(\omega-\Omega_{3,2})^{2}} + \frac{2\Gamma \sin^{2}(\alpha/2)\sin^{4}(\beta/2)}{\Gamma^{2}+(\omega-\Omega_{4,2})^{2}},\nonumber\\ 
\end{eqnarray}
with $\Omega_{k,1/2} \equiv E_{k}-E_{1/2}$ ($k=3,4$) and the $E_{i}$'s being the eigenvalues associated with the total Hamiltonian (the superscripts distinguish between TLS1 or TLS2). This result is valid to all orders of mutual TLS's coupling $g$, thereby giving rise to four components in the spectrum. Nevertheless, observation of numerical results reveals that, for the parameter regime and strain domain used in this discussion, $\cos^{2}(\alpha/2) \sim 1$, and the last two terms can be negligible since $\sin^{2}(\alpha/2) \sim 0$. Thus, the foregoing result contracts to
\begin{equation}
S_{eg}^{(1)}(\omega) \approx \frac{2\Gamma \cos^{4}(\beta/2)}{\Gamma^{2}+(\omega-\Omega_{3,1})^{2}} + \frac{2\Gamma \sin^{4} (\beta/2)}{\Gamma^{2}+(\omega-\Omega_{4,1})^{2}}.
\end{equation}
These two prevailing spectral components are also predicted by our effective model (\ref{eq:heff}), leading to essentially the  same result. Indeed, following the procedure outlined above, and under the same initial conditions but with a \textcolor{black}{much more} simplified time evolution operator $\exp(-iH_{\rm eff}t)$, yields the effective TLS1's  physical  spectrum:  
\begin{equation}
S_{eg}^{(\rm eff,1)}(\omega) = \frac{2 \Gamma \cos^{4}(\beta/2)}{\Gamma^{2}+(\omega-\bar{\Omega}_{3,1})^{2}} + \frac{2\Gamma \sin^{4} (\beta/2)}{\Gamma^{2}+(\omega-\bar{\Omega}_{4,1})^{2}},
\label{eq:effspectrum1}
\end{equation}
where $\bar{\Omega}_{k,1} = E^{\rm eff}_k-E^{\rm eff}_1$, with $E^{\rm eff}_k$ ($k=3,4$) being given, correspondingly, by the set of effective eigenvalues (\ref{eq:eff1})--(\ref{eq:eff4}). \textcolor{black}{As a function of $\omega$, each of the two terms in Eq.~(\ref{eq:effspectrum1}) represents a Lorentzian peak of full width at half maximum (FWHM) given by $2\Gamma$ and centered at  $\bar\Omega_{k,1}$.}
Fig.~\ref{fig:scheme4} displays the graphical result of the two Lorentzian line-shape functions (\ref{eq:effspectrum1}) for the same system parameters as in Fig.~\ref{fig:Elevels}. The clear evidence of the strain-dependent manifestation of mutual defect interplay is two-fold: $i)$ We observe a clear-cut S-shaped frequency shift that stems from the longitudinal coupling, $g_{\parallel}$, taking part in the interaction term $\propto \tilde{\sigma}_{z}^{(1)}\tilde{\sigma}_{z}^{(2)}$; $ii)$ Two avoided level crossings come about at specific resonant positions, where the  TLSs' energy exchange, caused by the interaction $\propto \tilde{\sigma}_{+}^{(1)}\tilde{\sigma}_{-}^{(2)}+\textrm{h.c.}$, weighted by the transversal coupling, $g_{\perp}$, is maximum  (the right one is zoomed in on the right panel of the figure located at $\sim 7.8$ V). This spectral feature is consistent with the fact that the TLS1 is excited, and it is closely connected with experimental observations through swap-spectroscopy already reported in AIO$_{x}$, see Fig.~3 of~\cite{Lisenfeld2015}. Unlike the energy spectrum representation depicted in Fig.~\ref{fig:Elevels}, the outcome in Fig.~\ref{fig:scheme4} can be construed as a proxy for the defect's strain-dependent power spectrum. Notice that the TLS2's hyperbolic dependence~(\ref{eq:splitting}), depicted by the gray dashed line in Fig.~\ref{fig:Elevels}, is invisible in Fig.~\ref{fig:scheme4}. This is because, except for the avoided level crossings, the initial excitation of the coupled system resides in the TLS1.

\begin{figure}[t]
\includegraphics[scale=0.4]{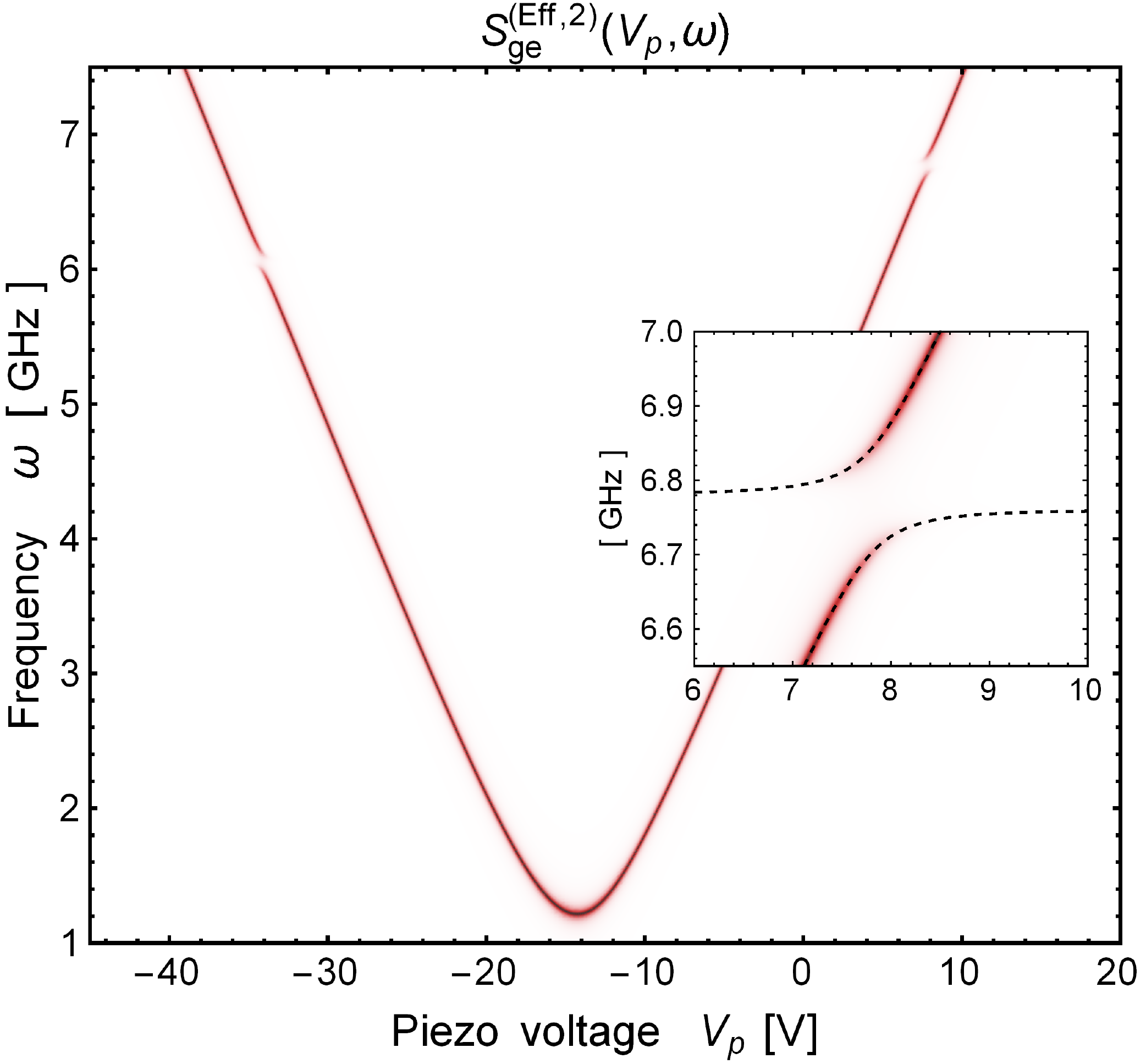}
\caption{Strain-dependent spectrum of TLS2 for the initial state $|g,e\rangle$, Eq.~(\ref{eq:effspec201}). The inset spotlight the spectral profiles in the neighborhood of defect resonances at $\sim 7.8$ V. The system parameters are the same as in Fig.~\ref{fig:Elevels} with $\Gamma=0.021$GHz.} 
\label{fig:scheme5}
\end{figure}

\begin{figure}[t]
\includegraphics[scale=0.4]{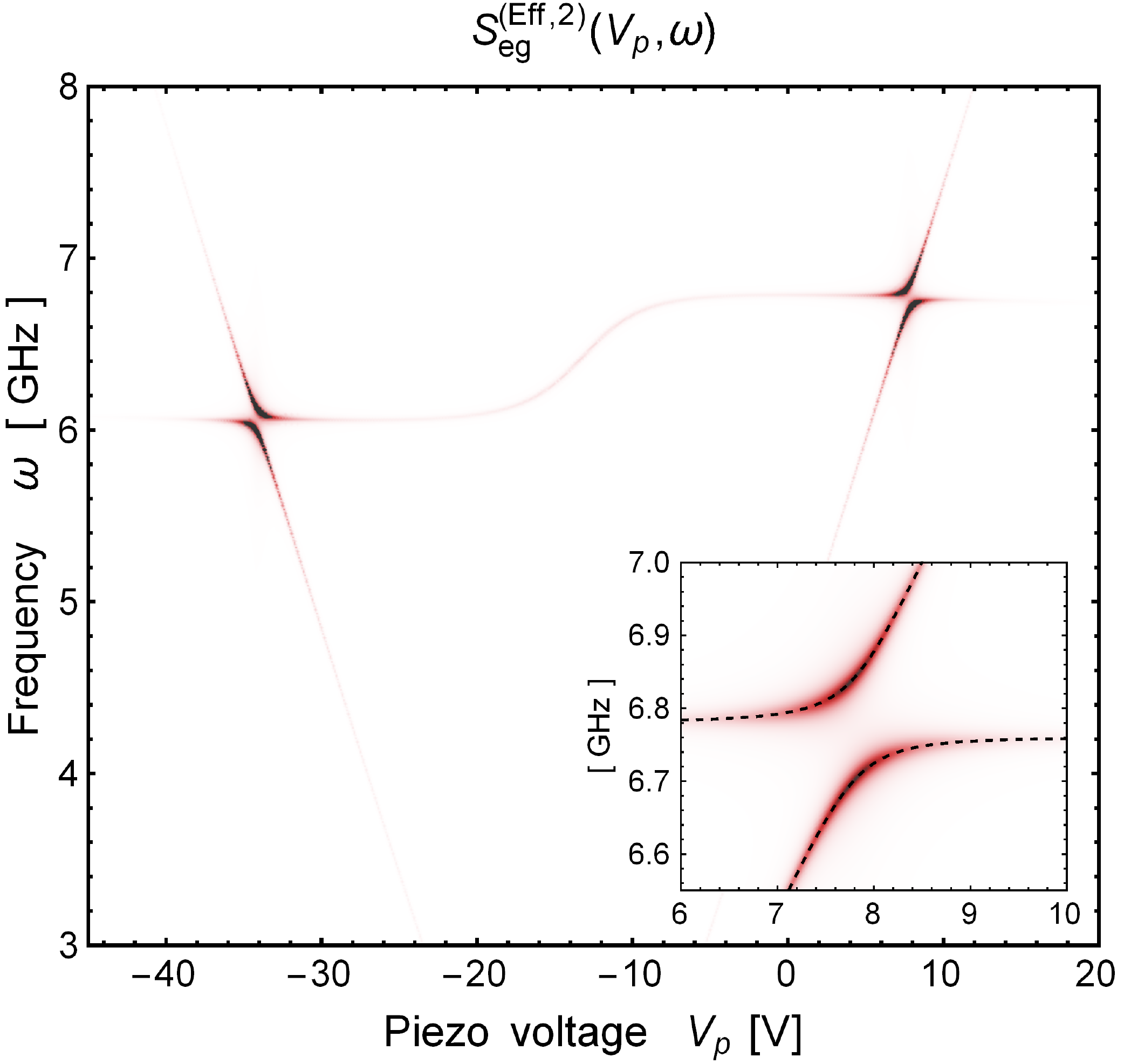} 
\caption{Strain-dependent spectrum of TLS2 for the initial state $|e,g\rangle$, Eq.~(\ref{eq:effspec210}). The inset spotlight the spectral profiles in the neighborhood of defect resonances at $\sim 7.8$ V. The system parameters are the same as in Fig.~\ref{fig:Elevels} with $\Gamma=0.01$GHz.} 
\label{fig:scheme6}
\end{figure}

Admitting the possibility of being able to implement TLS2 spectrum measurements, we show in Figs.~\ref{fig:scheme5} and~\ref{fig:scheme6} the corresponding predictions as to whether the initial state of system is the $|g,e\rangle=|g\rangle\otimes|e\rangle$ or the $|e,g\rangle$ state; that is, TLS1 is in its ground state whereas TLS2 is excited, or viceversa. So, the respective spectra of TLS2 are given by
%
\begin{subequations}
\begin{eqnarray}
S_{ge}^{(\textrm{eff},2)}(\omega)  & = &  \frac{2\Gamma \sin^{4} (\beta/2)}{\Gamma^{2}+(\omega-\bar{\Omega}_{3,1})^{2}} + \frac{2\textcolor{black}{\Gamma} \cos^{4}(\beta/2)}{\Gamma^{2}+(\omega-\bar{\Omega}_{4,1})^{2}},\qquad \label{eq:effspec201} \\
S_{eg}^{(\textrm{eff},2)}(\omega) & = & \frac{\Gamma \sin^{2}(\beta)/2}{\Gamma^{2}+(\omega-\bar{\Omega}_{3,1})^{2}}+\frac{\Gamma \sin^{2}(\beta)/2}{\Gamma^{2}+(\omega-\bar{\Omega}_{4,1})^{2}},
\qquad \label{eq:effspec210}
\end{eqnarray}
\end{subequations}
\textcolor{black}{where $\sin^2(\beta)=g_\perp^2/[\bar\omega_1-\bar\omega_2)^2+g_\perp^2]$.}

Depicting the TLS2's spectrum over a wider spectral detuning, we observe, in Fig.~\ref{fig:scheme5}, the well-known hyperbolic evolution~(\ref{eq:splitting}), with the symmetry point being discernible and located at approximately  $-13$ V. Again, the interplay between TLSs manifests itself in two avoided level crossings (being located at the same positions as in the previous case); one of them, at $\sim 7.8$ V, being zoomed in on the inset of the figure. In this case, one cannot see the characteristic S-shaped signal of TLS1 because the excitation is essentially in the TLS2. Fig.~\ref{fig:scheme6}, on the the other hand, shows the converse case in which the TLS1 is excited. In this case, we see that the TLS2's spectrum becomes conspicuous insofar as the spectral evolution between defects approach the avoided-crossing-level positions, where the resonant energy exchange is maximum. We note in passing that the feasibility of observing similar spectral features has recently been demonstrated, experimentally, via novel quantum sensors to detect TLS-TLS resonances in dielectrics \cite{Bilmes2021}, using both applied mechanical strain and DC electric fields for characterizing their spectroscopy.  For instance, Fig.~3 of~\cite{Bilmes2021} clearly shows segmented hyperbolic traces, as well as Supplementary Fig.~1(c) of~\cite{Lisenfeld2015}.

\begin{figure}
\begin{center}
\includegraphics[scale=0.4]{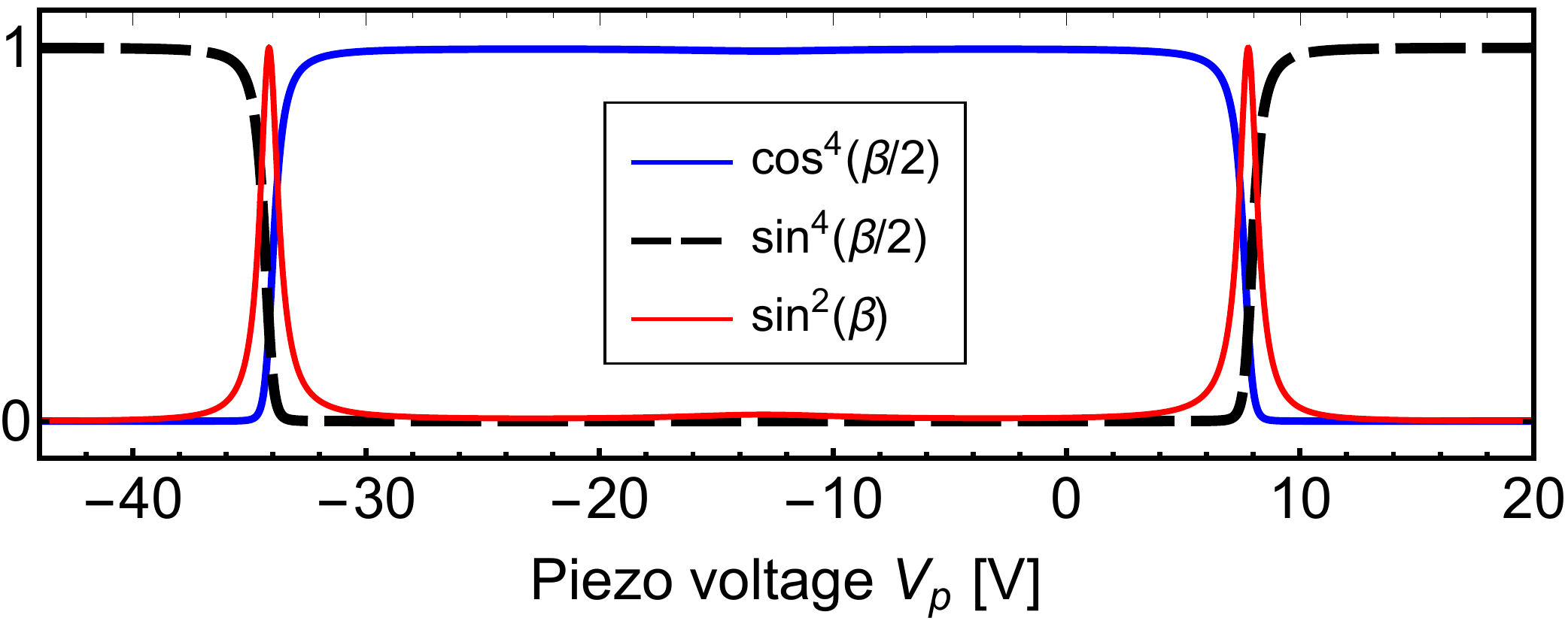}
\caption{ \textcolor{black}{ Weights associated to each Lorentzian function that shape the overall profile of the physical spectrum given in Eqs.~(\ref{eq:effspectrum1}), (\ref{eq:effspec201}) and (\ref{eq:effspec210}).}}
\label{amplitudes}
\end{center}
\end{figure}

\textcolor{black}{
For better understanding of the aforementioned behavior, we now discuss the role played by the weighing terms, $\cos^4(\beta/2)$ and $\sin^4(\beta/4)$, in Eq.~(\ref{eq:effspectrum1}), that shape the overall profile of the spectrum as a function of the piezo voltage. For instance, we see from Fig.~\ref{amplitudes} that the $\cos^4(\beta/2)$ and $\sin^4(\beta/4)$ terms are complementary to each other in Eq.~(\ref{eq:effspectrum1}), in the sense that they filter the spectral signal far from resonance, the combination of them giving rise to the Lorentzian-type spectrum: the first term of the spectrum, weighted by $\cos^4(\beta/2)$ (blue line), is essentially responsible for the S-shape signal (Fig.~\ref{fig:scheme4}) to appear in the region of about $-34 \ \textrm{V}  \le V_{p} \le 8 \ \textrm{V}$, whereas the complement trace of the spectrum, outside the aforesaid region, is weighted by the $\sin^4(\beta/4)$ term (dashed-black line) for $V_{p} < - 34$ V and $V_{p} > 8$ V. The contribution of the $\sin^2(\beta)$ term of Eq.~(\ref{eq:effspec210}) (red line), on the other hand, reflects only the strong interaction between TLSs near resonance, see Fig.~\ref{fig:scheme6}.
}

\textcolor{black}{
So far, we have concentrated on the case where the initial state contains a single excitation, $|e,g\rangle$ or $|g,e\rangle$. However, interesting results may arise if the system of two coupled TLS starts in the fully excited state $|e,e\rangle$. For the sake of simplicity, we will restrict ourselves to computing the spectral response of TLS1 only. It is easy to show that the corresponding EW spectrum, when using the initial state $|e,e\rangle$, is
}
\textcolor{black}{
\begin{equation}
S_{ee}^{(\rm eff,1)}(\omega) = \frac{2 \Gamma \sin^{4}(\beta/2)}{\Gamma^{2}+(\omega-\bar{\Omega}_{2,3})^{2}} + \frac{2\Gamma \cos^{4} (\beta/2)}{\Gamma^{2}+(\omega-\bar{\Omega}_{2,4})^{2}},
\end{equation} }
\textcolor{black}{where $\bar\Omega_{2,k}=E_2^{\rm eff}-E_k^{\rm eff}$. As a function of the applied piezo voltage, $S_{ee}^{(\rm eff,1)}(\omega)$ behaves like an inverted S-shaped signal (\reflectbox{S}), i.e., a signal similar to that of Fig.~\ref{fig:scheme4} but with a left avoided level crossing located in a frequency region that is larger than the right one. This result is consistent with the experimental measurements obtained in~\cite{Lisenfeld2015} using a two-photon swap spectroscopy technique, which brought the system to the state $|e,e\rangle$. However, in~\cite{Lisenfeld2015} only the right avoided level crossing was examined. In contrast, here we predict an entirely inverted S-shaped signal (\reflectbox{S}) for the $|e,e\rangle$ initial state.}

Focusing our attention on the insets of Figs.~\ref{fig:scheme4}, \ref{fig:scheme5}, and \ref{fig:scheme6}\textcolor{black}{, where $g_\perp\sim 0.14$GHz.} The transition energies, denoted by the black dashed lines, are the same in all of them. However, the corresponding spectral response differs significantly. For example, the EW spectrum in the inset of Fig.~\ref{fig:scheme4} tends to follow the TLS1's (horizontal) frequency, while for the inset of Fig.~\ref{fig:scheme5}, it follows the TLS2's (hyperbolic) frequency. Quite different is the behavior presented in Fig.~\ref{fig:scheme6}, where the spectral response near the avoided level crossing looks almost symmetric.
\textcolor{black}{The conspicuousness of the avoided level crossing (known as Rabi splitting) in such spectral responses confirms the fact that the TLS's interplay falls into the strong-coupling regime, i.e., the regime where the coupling strength, $g_\perp$, exceeds any individual decoherence rate of both TLS1 and TLS2; for all these cases, $g_\perp\gg \Gamma$. Indeed, in an experimental setting, the FWHM of each Lorentzian in the Rabi splitting can be considered proportional to the system's dissipation and decoherence rates. Therefore, we may incorporate some dissipation effects in the TLS spectra by tailoring the value of $\Gamma$.}

\begin{figure}[h]
\centering
\includegraphics[scale=0.38]{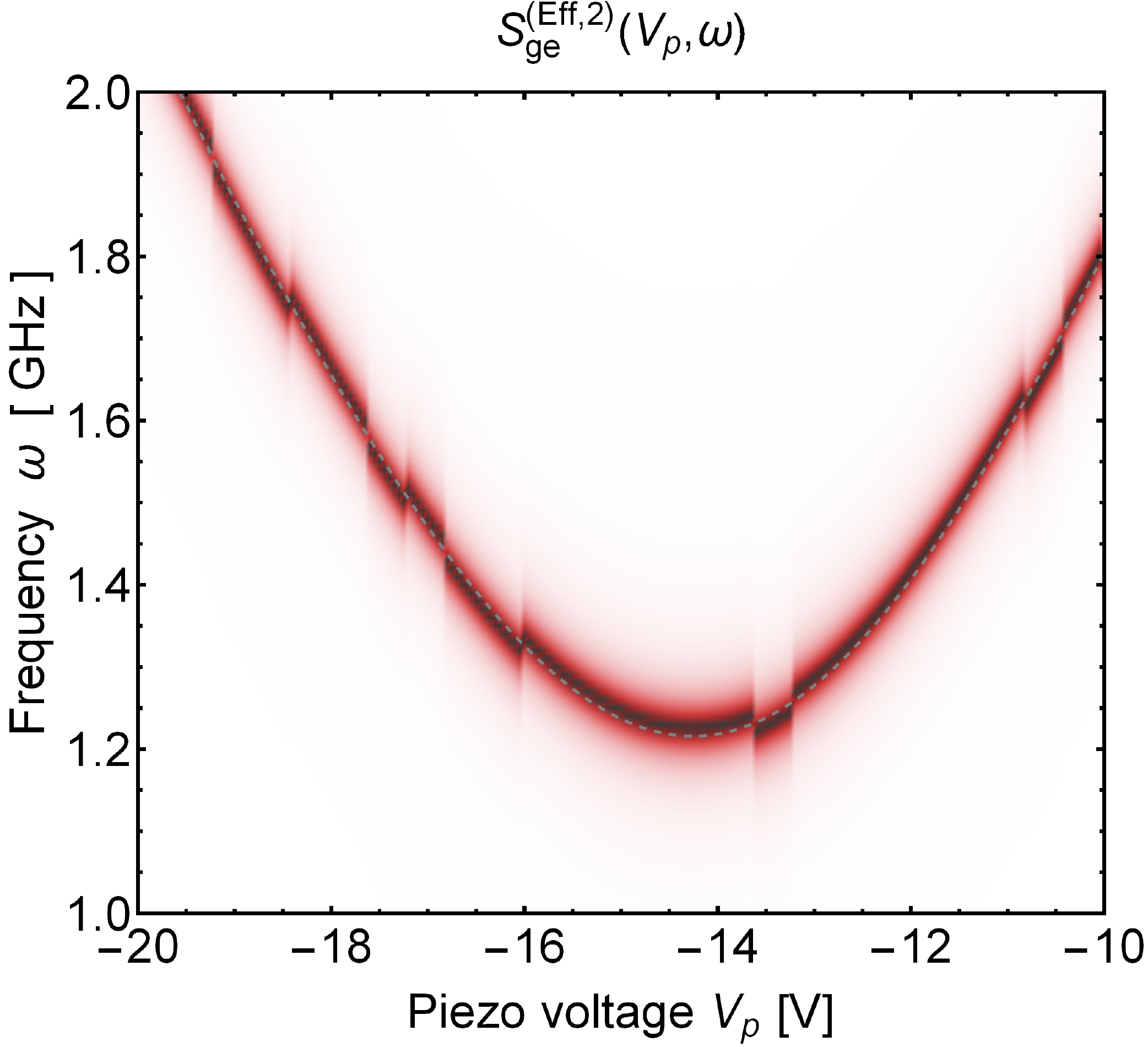}
\caption{\textcolor{black}{Typical telegraphic signal of the TLS2 resonance frequency. This behavior is due to the interaction of TLS2 with a dominant low-frequency TLS (thermal fluctuator). The system parameters are the same as in Fig.~\ref{fig:scheme5}, but for a plotted region near the symmetry point of TLS2.}}
\label{frequency_switching}
\end{figure}

\textcolor{black}{TLS can interact not only with their hosting device but also with their own environment~\cite{Muller_2019}. The coupling to phonon modes is the canonical source of dissipation for TLS~\cite{esquinazi2013tunneling}. Besides, the relaxation of TLS can also be due to their interaction with conduction electrons~\cite{Metallic_Glasses} and Bogoliubov quasiparticles~\cite{ElectronicDecoherence}. However, their coupling to other low-frequency TLS, known as thermal fluctuators, is of utmost importance because it causes incoherent state switching and spectral diffusion in the TLS~\cite{PRB_spectral_diffusion_16}. In several experiments, this decoherence mechanism produces spectroscopic signals of the TLS resonance frequency displaying telegraphic fluctuations~\cite{tunneling_fluctuators,Lisenfeld2015,Muller_2019}.  Remarkably, with the analytical spectra here obtained, we can easily reproduce such effect by adding random fluctuations of a few MHz to $\bar\Omega_{k,1}$.
We show this in Fig.~\ref{frequency_switching}, where the resonance frequency of the TLS2 (gray dashed-line) displays telegraphic fluctuations of $\sim\pm$10MHz~\cite{tunneling_fluctuators}. For comparison with a similar but experimental telegraphic signal, see  Supplementary Fig.~{1.c} of~\cite{Lisenfeld2015}.
}

\textcolor{black}{We note in passing} that, in the context of  strain tuning experiments~\cite{Grabovskij232}, the visibility of the spectral lines also depends on the coupling between each TLS with the superconducting qubit used as a detector. Thus, in principle, we could refine our results by considering such coupling. To do that, we need to include in~(\ref{eq:ham1}) the interaction Hamiltonian given in the Supplementary Note 3 of~\cite{Lisenfeld2015}; however, this goes beyond the central goal of this work.

\textcolor{black}{
Overall, we consider that our propounded algebraic procedure (including the effective Hamiltonian model and results that follow from it) is efficient in terms of its algebraic simplicity and its capability to offer a reliable insight about an experimentalist is observing or might expect to observe under certain conditions, like those encounter in the weak or strong coupling regime. Furthermore, the theory incorporates parameter settings that can be adjusted at will (for instance, the applied piezo voltage) and allows to explore and identify possible effects of the strain upon the measurements, in terms of the semiquantitative spectrum provided by the theoretical model. Our findings reveal that one is, in principle, capable of identifying, on the basis of the comparison between the theoretical and experimental outcomes, the initial state of the system and the coupling regime of TLS-TLS interplay, as to, for instance, whether or not the counter-rotating terms of the interaction Hamiltonian should be kept; we determined that this is not the case. Instead, the model was recast to a more straightforward and effective Hamiltonian that only retains energy-preserving rotating terms, which makes, in the strong coupling interaction case, the interpretation of the results simpler based upon the well-known rotating-wave approximation viewpoint.
}


\section{Conclusions}\label{Conclusions}

We have proposed a simple description of spectral signatures of strongly interacting TLS located inside the tunnel barrier of JJs of superconducting qubits. The proposal is based upon the standard atomic tunneling model to analyze the interplay of coupled defects viewed as dipole-dipole interacting TLS. In the strong coupling regime, analytical expressions for the energy and power spectrum of two coherently interacting TLS are obtained via the full Lisenfeld model. Using a small rotation~\cite{Klimov_Sanchez}, we show that, for moderate coupling, a further simplified version of it, an effective model~(\ref{eq:heff}), turns out to offer an accurate replication of the outcomes derived from spectroscopic experimental implementations~\cite{Lisenfeld2015}. The analytical description of coupled defects relies on the employment of the physical spectrum of Eberly-W\'odkiewicz by variation of their asymmetry energy $\varepsilon$, which depends linearly upon mechanical strain stemming from bending the chip circuit. The strain-tuning spectroscopy thus characterized can reveal the experimentally observable S-shaped frequency shift, and the expected level repulsion manifested as avoided level crossings, ascribable to strong defects' mutual dipole coupling, in addition to allowing us to trace the probability of the defect to be in a given initial state. Hopefully, experimentalists may adopt this approach\textcolor{black}{, valid for any initial state of the composite system,} as a supplemental tool to boost current studies of individual TLS. For instance, the EW spectrum could envision different types of Hamiltonians, describing unexplored spectral signatures of coupled TLS within disordered solids through strain-tuning spectroscopy. Furthermore, extractable information from this kind of spectroscopic explorations has become of particular relevance, in the context of quantum computers' performance, with a view to creating superconducting qubits of optimal functionality made up of low-loss insulators.
\medskip

\textbf{Disclosures.} The authors declare no conflicts of interest.

\textbf{Data availability statement.} All data that support the findings of this study are included within the article.


\begin{thebibliography}{50}%
\makeatletter
\providecommand \@ifxundefined [1]{%
 \@ifx{#1\undefined}
}%
\providecommand \@ifnum [1]{%
 \ifnum #1\expandafter \@firstoftwo
 \else \expandafter \@secondoftwo
 \fi
}%
\providecommand \@ifx [1]{%
 \ifx #1\expandafter \@firstoftwo
 \else \expandafter \@secondoftwo
 \fi
}%
\providecommand \natexlab [1]{#1}%
\providecommand \enquote  [1]{``#1''}%
\providecommand \bibnamefont  [1]{#1}%
\providecommand \bibfnamefont [1]{#1}%
\providecommand \citenamefont [1]{#1}%
\providecommand \href@noop [0]{\@secondoftwo}%
\providecommand \href [0]{\begingroup \@sanitize@url \@href}%
\providecommand \@href[1]{\@@startlink{#1}\@@href}%
\providecommand \@@href[1]{\endgroup#1\@@endlink}%
\providecommand \@sanitize@url [0]{\catcode `\\12\catcode `\$12\catcode
  `\&12\catcode `\#12\catcode `\^12\catcode `\_12\catcode `\%12\relax}%
\providecommand \@@startlink[1]{}%
\providecommand \@@endlink[0]{}%
\providecommand \url  [0]{\begingroup\@sanitize@url \@url }%
\providecommand \@url [1]{\endgroup\@href {#1}{\urlprefix }}%
\providecommand \urlprefix  [0]{URL }%
\providecommand \Eprint [0]{\href }%
\providecommand \doibase [0]{http://dx.doi.org/}%
\providecommand \selectlanguage [0]{\@gobble}%
\providecommand \bibinfo  [0]{\@secondoftwo}%
\providecommand \bibfield  [0]{\@secondoftwo}%
\providecommand \translation [1]{[#1]}%
\providecommand \BibitemOpen [0]{}%
\providecommand \bibitemStop [0]{}%
\providecommand \bibitemNoStop [0]{.\EOS\space}%
\providecommand \EOS [0]{\spacefactor3000\relax}%
\providecommand \BibitemShut  [1]{\csname bibitem#1\endcsname}%
\let\auto@bib@innerbib\@empty
\bibitem [{\citenamefont {Wallraff}\ \emph {et~al.}(2004)\citenamefont
  {Wallraff}, \citenamefont {Schuster}, \citenamefont {Blais}, \citenamefont
  {Frunzio}, \citenamefont {Huang}, \citenamefont {Majer}, \citenamefont
  {Kumar}, \citenamefont {Girvin},\ and\ \citenamefont
  {Schoelkopf}}]{Wallraff2004}%
  \BibitemOpen
  \bibfield  {author} {\bibinfo {author} {\bibfnamefont {A.}~\bibnamefont
  {Wallraff}}, \bibinfo {author} {\bibfnamefont {D.~I.}\ \bibnamefont
  {Schuster}}, \bibinfo {author} {\bibfnamefont {A.}~\bibnamefont {Blais}},
  \bibinfo {author} {\bibfnamefont {L.}~\bibnamefont {Frunzio}}, \bibinfo
  {author} {\bibfnamefont {R.-.~S.}\ \bibnamefont {Huang}}, \bibinfo {author}
  {\bibfnamefont {J.}~\bibnamefont {Majer}}, \bibinfo {author} {\bibfnamefont
  {S.}~\bibnamefont {Kumar}}, \bibinfo {author} {\bibfnamefont {S.~M.}\
  \bibnamefont {Girvin}}, \ and\ \bibinfo {author} {\bibfnamefont {R.~J.}\
  \bibnamefont {Schoelkopf}},\ }\bibfield  {title} {\enquote {\bibinfo {title}
  {Strong coupling of a single photon to a superconducting qubit using circuit
  quantum electrodynamics},}\ }\href {\doibase 10.1038/nature02851} {\bibfield
  {journal} {\bibinfo  {journal} {Nature}\ }\textbf {\bibinfo {volume} {431}},\
  \bibinfo {pages} {162--167} (\bibinfo {year} {2004})}\BibitemShut {NoStop}%
\bibitem [{\citenamefont {Arute~{\it et al}}(2019)}]{Google_2019}%
  \BibitemOpen
  \bibfield  {author} {\bibinfo {author} {\bibfnamefont {Frank}\ \bibnamefont
  {Arute~{\it et al}}},\ }\bibfield  {title} {\enquote {\bibinfo {title}
  {Quantum supremacy using a programmable superconducting processor},}\ }\href
  {\doibase 10.1038/s41586-019-1666-5} {\bibfield  {journal} {\bibinfo
  {journal} {Nature}\ }\textbf {\bibinfo {volume} {574}},\ \bibinfo {pages}
  {505--510} (\bibinfo {year} {2019})}\BibitemShut {NoStop}%
\bibitem [{\citenamefont {Blais}\ \emph {et~al.}(2021)\citenamefont {Blais},
  \citenamefont {Grimsmo}, \citenamefont {Girvin},\ and\ \citenamefont
  {Wallraff}}]{blais2020circuit}%
  \BibitemOpen
  \bibfield  {author} {\bibinfo {author} {\bibfnamefont {Alexandre}\
  \bibnamefont {Blais}}, \bibinfo {author} {\bibfnamefont {Arne~L.}\
  \bibnamefont {Grimsmo}}, \bibinfo {author} {\bibfnamefont {S.~M.}\
  \bibnamefont {Girvin}}, \ and\ \bibinfo {author} {\bibfnamefont {Andreas}\
  \bibnamefont {Wallraff}},\ }\bibfield  {title} {\enquote {\bibinfo {title}
  {Circuit quantum electrodynamics},}\ }\href {\doibase
  10.1103/RevModPhys.93.025005} {\bibfield  {journal} {\bibinfo  {journal}
  {Rev. Mod. Phys.}\ }\textbf {\bibinfo {volume} {93}},\ \bibinfo {pages}
  {025005} (\bibinfo {year} {2021})}\BibitemShut {NoStop}%
\bibitem [{\citenamefont {Somoroff}\ \emph {et~al.}(2021)\citenamefont
  {Somoroff}, \citenamefont {Ficheux}, \citenamefont {Mencia}, \citenamefont
  {Xiong}, \citenamefont {Kuzmin},\ and\ \citenamefont
  {Manucharyan}}]{somoroff2021millisecond}%
  \BibitemOpen
  \bibfield  {author} {\bibinfo {author} {\bibfnamefont {Aaron}\ \bibnamefont
  {Somoroff}}, \bibinfo {author} {\bibfnamefont {Quentin}\ \bibnamefont
  {Ficheux}}, \bibinfo {author} {\bibfnamefont {Raymond~A.}\ \bibnamefont
  {Mencia}}, \bibinfo {author} {\bibfnamefont {Haonan}\ \bibnamefont {Xiong}},
  \bibinfo {author} {\bibfnamefont {Roman~V.}\ \bibnamefont {Kuzmin}}, \ and\
  \bibinfo {author} {\bibfnamefont {Vladimir~E.}\ \bibnamefont {Manucharyan}},\
  }\href@noop {} {\enquote {\bibinfo {title} {Millisecond coherence in a
  superconducting qubit},}\ } (\bibinfo {year} {2021}),\ \Eprint
  {http://arxiv.org/abs/2103.08578} {arXiv:2103.08578 [quant-ph]} \BibitemShut
  {NoStop}%
\bibitem [{\citenamefont {Martinis}(2021)}]{Martinis2021}%
  \BibitemOpen
  \bibfield  {author} {\bibinfo {author} {\bibfnamefont {John~M.}\ \bibnamefont
  {Martinis}},\ }\bibfield  {title} {\enquote {\bibinfo {title} {Saving
  superconducting quantum processors from decay and correlated errors generated
  by gamma and cosmic rays},}\ }\href {\doibase 10.1038/s41534-021-00431-0}
  {\bibfield  {journal} {\bibinfo  {journal} {npj Quantum Information}\
  }\textbf {\bibinfo {volume} {7}},\ \bibinfo {pages} {90} (\bibinfo {year}
  {2021})}\BibitemShut {NoStop}%
\bibitem [{\citenamefont {Cardani~{\it et al}}(2020)}]{Cardani2020}%
  \BibitemOpen
  \bibfield  {author} {\bibinfo {author} {\bibfnamefont {L.}~\bibnamefont
  {Cardani~{\it et al}}},\ }\bibfield  {title} {\enquote {\bibinfo {title}
  {{DEMETRA}: Suppression of the relaxation induced by radioactivity in
  superconducting qubits},}\ }\href {\doibase 10.1007/s10909-019-02265-9}
  {\bibfield  {journal} {\bibinfo  {journal} {Journal of Low Temperature
  Physics}\ }\textbf {\bibinfo {volume} {199}},\ \bibinfo {pages} {475--481}
  (\bibinfo {year} {2020})}\BibitemShut {NoStop}%
\bibitem [{\citenamefont {Murray}(2021)}]{Murray2021}%
  \BibitemOpen
  \bibfield  {author} {\bibinfo {author} {\bibfnamefont {Conal~E.}\
  \bibnamefont {Murray}},\ }\bibfield  {title} {\enquote {\bibinfo {title}
  {Material matters in superconducting qubits},}\ }\href {\doibase
  https://doi.org/10.1016/j.mser.2021.100646} {\bibfield  {journal} {\bibinfo
  {journal} {Materials Science and Engineering: R: Reports}\ }\textbf {\bibinfo
  {volume} {146}},\ \bibinfo {pages} {100646} (\bibinfo {year}
  {2021})}\BibitemShut {NoStop}%
\bibitem [{\citenamefont {Simmonds}\ \emph {et~al.}(2004)\citenamefont
  {Simmonds}, \citenamefont {Lang}, \citenamefont {Hite}, \citenamefont {Nam},
  \citenamefont {Pappas},\ and\ \citenamefont {Martinis}}]{Martinis_PRL_2004}%
  \BibitemOpen
  \bibfield  {author} {\bibinfo {author} {\bibfnamefont {R.~W.}\ \bibnamefont
  {Simmonds}}, \bibinfo {author} {\bibfnamefont {K.~M.}\ \bibnamefont {Lang}},
  \bibinfo {author} {\bibfnamefont {D.~A.}\ \bibnamefont {Hite}}, \bibinfo
  {author} {\bibfnamefont {S.}~\bibnamefont {Nam}}, \bibinfo {author}
  {\bibfnamefont {D.~P.}\ \bibnamefont {Pappas}}, \ and\ \bibinfo {author}
  {\bibfnamefont {John~M.}\ \bibnamefont {Martinis}},\ }\bibfield  {title}
  {\enquote {\bibinfo {title} {Decoherence in {J}osephson phase qubits from
  junction resonators},}\ }\href {\doibase 10.1103/PhysRevLett.93.077003}
  {\bibfield  {journal} {\bibinfo  {journal} {Phys. Rev. Lett.}\ }\textbf
  {\bibinfo {volume} {93}},\ \bibinfo {pages} {077003} (\bibinfo {year}
  {2004})}\BibitemShut {NoStop}%
\bibitem [{\citenamefont {Martinis}\ \emph {et~al.}(2005)\citenamefont
  {Martinis}, \citenamefont {Cooper}, \citenamefont {McDermott}, \citenamefont
  {Steffen}, \citenamefont {Ansmann}, \citenamefont {Osborn}, \citenamefont
  {Cicak}, \citenamefont {Oh}, \citenamefont {Pappas}, \citenamefont
  {Simmonds},\ and\ \citenamefont {Yu}}]{Martinis_PRL_2005}%
  \BibitemOpen
  \bibfield  {author} {\bibinfo {author} {\bibfnamefont {John~M.}\ \bibnamefont
  {Martinis}}, \bibinfo {author} {\bibfnamefont {K.~B.}\ \bibnamefont
  {Cooper}}, \bibinfo {author} {\bibfnamefont {R.}~\bibnamefont {McDermott}},
  \bibinfo {author} {\bibfnamefont {Matthias}\ \bibnamefont {Steffen}},
  \bibinfo {author} {\bibfnamefont {Markus}\ \bibnamefont {Ansmann}}, \bibinfo
  {author} {\bibfnamefont {K.~D.}\ \bibnamefont {Osborn}}, \bibinfo {author}
  {\bibfnamefont {K.}~\bibnamefont {Cicak}}, \bibinfo {author} {\bibfnamefont
  {Seongshik}\ \bibnamefont {Oh}}, \bibinfo {author} {\bibfnamefont {D.~P.}\
  \bibnamefont {Pappas}}, \bibinfo {author} {\bibfnamefont {R.~W.}\
  \bibnamefont {Simmonds}}, \ and\ \bibinfo {author} {\bibfnamefont {Clare~C.}\
  \bibnamefont {Yu}},\ }\bibfield  {title} {\enquote {\bibinfo {title}
  {Decoherence in {J}osephson qubits from dielectric loss},}\ }\href {\doibase
  10.1103/PhysRevLett.95.210503} {\bibfield  {journal} {\bibinfo  {journal}
  {Phys. Rev. Lett.}\ }\textbf {\bibinfo {volume} {95}},\ \bibinfo {pages}
  {210503} (\bibinfo {year} {2005})}\BibitemShut {NoStop}%
\bibitem [{\citenamefont {Burin}\ \emph {et~al.}(2014)\citenamefont {Burin},
  \citenamefont {Maksymov},\ and\ \citenamefont {Osborn}}]{Burin_2014}%
  \BibitemOpen
  \bibfield  {author} {\bibinfo {author} {\bibfnamefont {Alexander~L}\
  \bibnamefont {Burin}}, \bibinfo {author} {\bibfnamefont {Andrii~O}\
  \bibnamefont {Maksymov}}, \ and\ \bibinfo {author} {\bibfnamefont {Kevin~D}\
  \bibnamefont {Osborn}},\ }\bibfield  {title} {\enquote {\bibinfo {title}
  {Quantum coherent manipulation of two-level systems in superconducting
  circuits},}\ }\href {\doibase 10.1088/0953-2048/27/8/084001} {\bibfield
  {journal} {\bibinfo  {journal} {Superconductor Science and Technology}\
  }\textbf {\bibinfo {volume} {27}},\ \bibinfo {pages} {084001} (\bibinfo
  {year} {2014})}\BibitemShut {NoStop}%
\bibitem [{\citenamefont {Lubchenko}\ and\ \citenamefont
  {Wolynes}(2007)}]{Lubchenko_2007}%
  \BibitemOpen
  \bibfield  {author} {\bibinfo {author} {\bibfnamefont {Vassiliy}\
  \bibnamefont {Lubchenko}}\ and\ \bibinfo {author} {\bibfnamefont {Peter~G.}\
  \bibnamefont {Wolynes}},\ }\enquote {\bibinfo {title} {The microscopic
  quantum theory of low temperature amorphous solids},}\ in\ \href {\doibase
  https://doi.org/10.1002/9780470175422.ch3} {\emph {\bibinfo {booktitle}
  {Advances in Chemical Physics}}}\ (\bibinfo  {publisher} {John Wiley \& Sons,
  Ltd},\ \bibinfo {year} {2007})\ Chap.~\bibinfo {chapter} {3}, pp.\ \bibinfo
  {pages} {95--206}\BibitemShut {NoStop}%
\bibitem [{\citenamefont {Leggett}\ and\ \citenamefont
  {Vural}(2013)}]{Leggett_2013}%
  \BibitemOpen
  \bibfield  {author} {\bibinfo {author} {\bibfnamefont {Anthony~J.}\
  \bibnamefont {Leggett}}\ and\ \bibinfo {author} {\bibfnamefont {Dervis~C.}\
  \bibnamefont {Vural}},\ }\bibfield  {title} {\enquote {\bibinfo {title}
  {Tunneling two-level systems model of the low-temperature properties of
  glasses: Are “{S}moking-{G}un” tests possible?}}\ }\href {\doibase
  10.1021/jp402222g} {\bibfield  {journal} {\bibinfo  {journal} {The Journal of
  Physical Chemistry B}\ }\textbf {\bibinfo {volume} {117}},\ \bibinfo {pages}
  {12966--12971} (\bibinfo {year} {2013})},\ \bibinfo {note} {pMID:
  23924397}\BibitemShut {NoStop}%
\bibitem [{\citenamefont {Esquinazi}(2013)}]{esquinazi2013tunneling}%
  \BibitemOpen
  \bibfield  {author} {\bibinfo {author} {\bibfnamefont {Pablo}\ \bibnamefont
  {Esquinazi}},\ }\href@noop {} {\emph {\bibinfo {title} {Tunneling systems in
  amorphous and crystalline solids}}}\ (\bibinfo  {publisher} {Springer Science
  \& Business Media},\ \bibinfo {year} {2013})\BibitemShut {NoStop}%
\bibitem [{\citenamefont {Yu}(2004)}]{Yu2004}%
  \BibitemOpen
  \bibfield  {author} {\bibinfo {author} {\bibfnamefont {Clare~C.}\
  \bibnamefont {Yu}},\ }\bibfield  {title} {\enquote {\bibinfo {title} {Why
  study noise due to two level systems: A suggestion for experimentalists},}\
  }\href {\doibase 10.1023/B:JOLT.0000049056.07100.85} {\bibfield  {journal}
  {\bibinfo  {journal} {Journal of Low Temperature Physics}\ }\textbf {\bibinfo
  {volume} {137}},\ \bibinfo {pages} {251--265} (\bibinfo {year}
  {2004})}\BibitemShut {NoStop}%
\bibitem [{\citenamefont {Lupa\ifmmode~\mbox{\c{s}}\else \c{s}\fi{}cu}\ \emph
  {et~al.}(2009)\citenamefont {Lupa\ifmmode~\mbox{\c{s}}\else \c{s}\fi{}cu},
  \citenamefont {Bertet}, \citenamefont {Driessen}, \citenamefont {Harmans},\
  and\ \citenamefont {Mooij}}]{Driessen_PRB_2009}%
  \BibitemOpen
  \bibfield  {author} {\bibinfo {author} {\bibfnamefont {A.}~\bibnamefont
  {Lupa\ifmmode~\mbox{\c{s}}\else \c{s}\fi{}cu}}, \bibinfo {author}
  {\bibfnamefont {P.}~\bibnamefont {Bertet}}, \bibinfo {author} {\bibfnamefont
  {E.~F.~C.}\ \bibnamefont {Driessen}}, \bibinfo {author} {\bibfnamefont {C.~J.
  P.~M.}\ \bibnamefont {Harmans}}, \ and\ \bibinfo {author} {\bibfnamefont
  {J.~E.}\ \bibnamefont {Mooij}},\ }\bibfield  {title} {\enquote {\bibinfo
  {title} {One- and two-photon spectroscopy of a flux qubit coupled to a
  microscopic defect},}\ }\href {\doibase 10.1103/PhysRevB.80.172506}
  {\bibfield  {journal} {\bibinfo  {journal} {Phys. Rev. B}\ }\textbf {\bibinfo
  {volume} {80}},\ \bibinfo {pages} {172506} (\bibinfo {year}
  {2009})}\BibitemShut {NoStop}%
\bibitem [{\citenamefont {Osman}\ \emph {et~al.}(2021)\citenamefont {Osman},
  \citenamefont {Simon}, \citenamefont {Bengtsson}, \citenamefont {Kosen},
  \citenamefont {Krantz}, \citenamefont {P.~Lozano}, \citenamefont
  {Scigliuzzo}, \citenamefont {Delsing}, \citenamefont {Bylander},\ and\
  \citenamefont {Fadavi~Roudsari}}]{Osman_2021}%
  \BibitemOpen
  \bibfield  {author} {\bibinfo {author} {\bibfnamefont {A.}~\bibnamefont
  {Osman}}, \bibinfo {author} {\bibfnamefont {J.}~\bibnamefont {Simon}},
  \bibinfo {author} {\bibfnamefont {A.}~\bibnamefont {Bengtsson}}, \bibinfo
  {author} {\bibfnamefont {S.}~\bibnamefont {Kosen}}, \bibinfo {author}
  {\bibfnamefont {P.}~\bibnamefont {Krantz}}, \bibinfo {author} {\bibfnamefont
  {D.}~\bibnamefont {P.~Lozano}}, \bibinfo {author} {\bibfnamefont
  {M.}~\bibnamefont {Scigliuzzo}}, \bibinfo {author} {\bibfnamefont
  {P.}~\bibnamefont {Delsing}}, \bibinfo {author} {\bibfnamefont {Jonas}\
  \bibnamefont {Bylander}}, \ and\ \bibinfo {author} {\bibfnamefont
  {A.}~\bibnamefont {Fadavi~Roudsari}},\ }\bibfield  {title} {\enquote
  {\bibinfo {title} {Simplified {J}osephson-junction fabrication process for
  reproducibly high-performance superconducting qubits},}\ }\href {\doibase
  10.1063/5.0037093} {\bibfield  {journal} {\bibinfo  {journal} {Applied
  Physics Letters}\ }\textbf {\bibinfo {volume} {118}},\ \bibinfo {pages}
  {064002} (\bibinfo {year} {2021})}\BibitemShut {NoStop}%
\bibitem [{\citenamefont {Bilmes}\ \emph
  {et~al.}(2021{\natexlab{a}})\citenamefont {Bilmes}, \citenamefont {Händel},
  \citenamefont {Volosheniuk}, \citenamefont {Ustinov},\ and\ \citenamefont
  {Lisenfeld}}]{In_situ_2021}%
  \BibitemOpen
  \bibfield  {author} {\bibinfo {author} {\bibfnamefont {Alexander}\
  \bibnamefont {Bilmes}}, \bibinfo {author} {\bibfnamefont {Alexander~K}\
  \bibnamefont {Händel}}, \bibinfo {author} {\bibfnamefont {Serhii}\
  \bibnamefont {Volosheniuk}}, \bibinfo {author} {\bibfnamefont {Alexey~V}\
  \bibnamefont {Ustinov}}, \ and\ \bibinfo {author} {\bibfnamefont {Jürgen}\
  \bibnamefont {Lisenfeld}},\ }\bibfield  {title} {\enquote {\bibinfo {title}
  {In-situ bandaged {J}osephson junctions for superconducting quantum
  processors},}\ }\href {\doibase 10.1088/1361-6668/ac2a6d} {\bibfield
  {journal} {\bibinfo  {journal} {Superconductor Science and Technology}\
  }\textbf {\bibinfo {volume} {34}},\ \bibinfo {pages} {125011} (\bibinfo
  {year} {2021}{\natexlab{a}})}\BibitemShut {NoStop}%
\bibitem [{\citenamefont {Earnest}\ \emph {et~al.}(2018)\citenamefont
  {Earnest}, \citenamefont {B{\'{e}}janin}, \citenamefont {McConkey},
  \citenamefont {Peters}, \citenamefont {Korinek}, \citenamefont {Yuan},\ and\
  \citenamefont {Mariantoni}}]{Earnest_2018}%
  \BibitemOpen
  \bibfield  {author} {\bibinfo {author} {\bibfnamefont {C~T}\ \bibnamefont
  {Earnest}}, \bibinfo {author} {\bibfnamefont {J~H}\ \bibnamefont
  {B{\'{e}}janin}}, \bibinfo {author} {\bibfnamefont {T~G}\ \bibnamefont
  {McConkey}}, \bibinfo {author} {\bibfnamefont {E~A}\ \bibnamefont {Peters}},
  \bibinfo {author} {\bibfnamefont {A}~\bibnamefont {Korinek}}, \bibinfo
  {author} {\bibfnamefont {H}~\bibnamefont {Yuan}}, \ and\ \bibinfo {author}
  {\bibfnamefont {M}~\bibnamefont {Mariantoni}},\ }\bibfield  {title} {\enquote
  {\bibinfo {title} {Substrate surface engineering for high-quality
  silicon/aluminum superconducting resonators},}\ }\href {\doibase
  10.1088/1361-6668/aae548} {\bibfield  {journal} {\bibinfo  {journal}
  {Superconductor Science and Technology}\ }\textbf {\bibinfo {volume} {31}},\
  \bibinfo {pages} {125013} (\bibinfo {year} {2018})}\BibitemShut {NoStop}%
\bibitem [{\citenamefont {Grabovskij}\ \emph {et~al.}(2012)\citenamefont
  {Grabovskij}, \citenamefont {Peichl}, \citenamefont {Lisenfeld},
  \citenamefont {Weiss},\ and\ \citenamefont {Ustinov}}]{Grabovskij232}%
  \BibitemOpen
  \bibfield  {author} {\bibinfo {author} {\bibfnamefont {Grigorij~J.}\
  \bibnamefont {Grabovskij}}, \bibinfo {author} {\bibfnamefont {Torben}\
  \bibnamefont {Peichl}}, \bibinfo {author} {\bibfnamefont {J{\"u}rgen}\
  \bibnamefont {Lisenfeld}}, \bibinfo {author} {\bibfnamefont {Georg}\
  \bibnamefont {Weiss}}, \ and\ \bibinfo {author} {\bibfnamefont {Alexey~V.}\
  \bibnamefont {Ustinov}},\ }\bibfield  {title} {\enquote {\bibinfo {title}
  {Strain tuning of individual atomic tunneling systems detected by a
  superconducting qubit},}\ }\href {\doibase 10.1126/science.1226487}
  {\bibfield  {journal} {\bibinfo  {journal} {Science}\ }\textbf {\bibinfo
  {volume} {338}},\ \bibinfo {pages} {232--234} (\bibinfo {year}
  {2012})}\BibitemShut {NoStop}%
\bibitem [{\citenamefont {Bilmes}\ \emph
  {et~al.}(2021{\natexlab{b}})\citenamefont {Bilmes}, \citenamefont
  {Volosheniuk}, \citenamefont {Brehm}, \citenamefont {Ustinov},\ and\
  \citenamefont {Lisenfeld}}]{Bilmes2021}%
  \BibitemOpen
  \bibfield  {author} {\bibinfo {author} {\bibfnamefont {Alexander}\
  \bibnamefont {Bilmes}}, \bibinfo {author} {\bibfnamefont {Serhii}\
  \bibnamefont {Volosheniuk}}, \bibinfo {author} {\bibfnamefont {Jan~David}\
  \bibnamefont {Brehm}}, \bibinfo {author} {\bibfnamefont {Alexey~V.}\
  \bibnamefont {Ustinov}}, \ and\ \bibinfo {author} {\bibfnamefont
  {J{\"u}rgen}\ \bibnamefont {Lisenfeld}},\ }\bibfield  {title} {\enquote
  {\bibinfo {title} {Quantum sensors for microscopic tunneling systems},}\
  }\href {\doibase 10.1038/s41534-020-00359-x} {\bibfield  {journal} {\bibinfo
  {journal} {npj Quantum Information}\ }\textbf {\bibinfo {volume} {7}},\
  \bibinfo {pages} {27} (\bibinfo {year} {2021}{\natexlab{b}})}\BibitemShut
  {NoStop}%
\bibitem [{\citenamefont {Müller}\ \emph {et~al.}(2019)\citenamefont
  {Müller}, \citenamefont {Cole},\ and\ \citenamefont
  {Lisenfeld}}]{Muller_2019}%
  \BibitemOpen
  \bibfield  {author} {\bibinfo {author} {\bibfnamefont {Clemens}\ \bibnamefont
  {Müller}}, \bibinfo {author} {\bibfnamefont {Jared~H}\ \bibnamefont {Cole}},
  \ and\ \bibinfo {author} {\bibfnamefont {Jürgen}\ \bibnamefont
  {Lisenfeld}},\ }\bibfield  {title} {\enquote {\bibinfo {title} {Towards
  understanding two-level-systems in amorphous solids: insights from quantum
  circuits},}\ }\href {\doibase 10.1088/1361-6633/ab3a7e} {\bibfield  {journal}
  {\bibinfo  {journal} {Reports on Progress in Physics}\ }\textbf {\bibinfo
  {volume} {82}},\ \bibinfo {pages} {124501} (\bibinfo {year}
  {2019})}\BibitemShut {NoStop}%
\bibitem [{\citenamefont {Zagoskin}\ \emph {et~al.}(2006)\citenamefont
  {Zagoskin}, \citenamefont {Ashhab}, \citenamefont {Johansson},\ and\
  \citenamefont {Nori}}]{Zagoskin_2006}%
  \BibitemOpen
  \bibfield  {author} {\bibinfo {author} {\bibfnamefont {A.~M.}\ \bibnamefont
  {Zagoskin}}, \bibinfo {author} {\bibfnamefont {S.}~\bibnamefont {Ashhab}},
  \bibinfo {author} {\bibfnamefont {J.~R.}\ \bibnamefont {Johansson}}, \ and\
  \bibinfo {author} {\bibfnamefont {Franco}\ \bibnamefont {Nori}},\ }\bibfield
  {title} {\enquote {\bibinfo {title} {Quantum two-level systems in {J}osephson
  junctions as naturally formed qubits},}\ }\href {\doibase
  10.1103/PhysRevLett.97.077001} {\bibfield  {journal} {\bibinfo  {journal}
  {Phys. Rev. Lett.}\ }\textbf {\bibinfo {volume} {97}},\ \bibinfo {pages}
  {077001} (\bibinfo {year} {2006})}\BibitemShut {NoStop}%
\bibitem [{\citenamefont {Neeley}\ \emph {et~al.}(2008)\citenamefont {Neeley},
  \citenamefont {Ansmann}, \citenamefont {Bialczak}, \citenamefont {Hofheinz},
  \citenamefont {Katz}, \citenamefont {Lucero}, \citenamefont {O'Connell},
  \citenamefont {Wang}, \citenamefont {Cleland},\ and\ \citenamefont
  {Martinis}}]{Neeley2008}%
  \BibitemOpen
  \bibfield  {author} {\bibinfo {author} {\bibfnamefont {Matthew}\ \bibnamefont
  {Neeley}}, \bibinfo {author} {\bibfnamefont {M.}~\bibnamefont {Ansmann}},
  \bibinfo {author} {\bibfnamefont {Radoslaw~C.}\ \bibnamefont {Bialczak}},
  \bibinfo {author} {\bibfnamefont {M.}~\bibnamefont {Hofheinz}}, \bibinfo
  {author} {\bibfnamefont {N.}~\bibnamefont {Katz}}, \bibinfo {author}
  {\bibfnamefont {Erik}\ \bibnamefont {Lucero}}, \bibinfo {author}
  {\bibfnamefont {A.}~\bibnamefont {O'Connell}}, \bibinfo {author}
  {\bibfnamefont {H.}~\bibnamefont {Wang}}, \bibinfo {author} {\bibfnamefont
  {A.~N.}\ \bibnamefont {Cleland}}, \ and\ \bibinfo {author} {\bibfnamefont
  {John~M.}\ \bibnamefont {Martinis}},\ }\bibfield  {title} {\enquote {\bibinfo
  {title} {Process tomography of quantum memory in a {J}osephson-phase qubit
  coupled to a two-level state},}\ }\href {\doibase 10.1038/nphys972}
  {\bibfield  {journal} {\bibinfo  {journal} {Nature Physics}\ }\textbf
  {\bibinfo {volume} {4}},\ \bibinfo {pages} {523--526} (\bibinfo {year}
  {2008})}\BibitemShut {NoStop}%
\bibitem [{\citenamefont {Burin}\ \emph {et~al.}(1998)\citenamefont {Burin},
  \citenamefont {Natelson}, \citenamefont {Osheroff},\ and\ \citenamefont
  {Kagan}}]{Burin1998}%
  \BibitemOpen
  \bibfield  {author} {\bibinfo {author} {\bibfnamefont {Alexander~L.}\
  \bibnamefont {Burin}}, \bibinfo {author} {\bibfnamefont {Douglas}\
  \bibnamefont {Natelson}}, \bibinfo {author} {\bibfnamefont {Douglas~D.}\
  \bibnamefont {Osheroff}}, \ and\ \bibinfo {author} {\bibfnamefont {Yuri}\
  \bibnamefont {Kagan}},\ }\enquote {\bibinfo {title} {Interactions between
  tunneling defects in amorphous solids},}\ in\ \href {\doibase
  10.1007/978-3-662-03695-2_5} {\emph {\bibinfo {booktitle} {Tunneling Systems
  in Amorphous and Crystalline Solids}}},\ \bibinfo {editor} {edited by\
  \bibinfo {editor} {\bibfnamefont {Pablo}\ \bibnamefont {Esquinazi}}}\
  (\bibinfo  {publisher} {Springer Berlin Heidelberg},\ \bibinfo {address}
  {Berlin, Heidelberg},\ \bibinfo {year} {1998})\ pp.\ \bibinfo {pages}
  {223--315}\BibitemShut {NoStop}%
\bibitem [{\citenamefont {Lisenfeld}\ \emph {et~al.}(2015)\citenamefont
  {Lisenfeld}, \citenamefont {Grabovskij}, \citenamefont {M{\"u}ller},
  \citenamefont {Cole}, \citenamefont {Weiss},\ and\ \citenamefont
  {Ustinov}}]{Lisenfeld2015}%
  \BibitemOpen
  \bibfield  {author} {\bibinfo {author} {\bibfnamefont {J{\"u}rgen}\
  \bibnamefont {Lisenfeld}}, \bibinfo {author} {\bibfnamefont {Grigorij~J.}\
  \bibnamefont {Grabovskij}}, \bibinfo {author} {\bibfnamefont {Clemens}\
  \bibnamefont {M{\"u}ller}}, \bibinfo {author} {\bibfnamefont {Jared~H.}\
  \bibnamefont {Cole}}, \bibinfo {author} {\bibfnamefont {Georg}\ \bibnamefont
  {Weiss}}, \ and\ \bibinfo {author} {\bibfnamefont {Alexey~V.}\ \bibnamefont
  {Ustinov}},\ }\bibfield  {title} {\enquote {\bibinfo {title} {Observation of
  directly interacting coherent two-level systems in an amorphous material},}\
  }\href {\doibase 10.1038/ncomms7182} {\bibfield  {journal} {\bibinfo
  {journal} {Nature Communications}\ }\textbf {\bibinfo {volume} {6}},\
  \bibinfo {pages} {6182} (\bibinfo {year} {2015})}\BibitemShut {NoStop}%
\bibitem [{\citenamefont {Lisenfeld}\ \emph {et~al.}(2019)\citenamefont
  {Lisenfeld}, \citenamefont {Bilmes}, \citenamefont {Megrant}, \citenamefont
  {Barends}, \citenamefont {Kelly}, \citenamefont {Klimov}, \citenamefont
  {Weiss}, \citenamefont {Martinis},\ and\ \citenamefont
  {Ustinov}}]{Lisenfeld2019}%
  \BibitemOpen
  \bibfield  {author} {\bibinfo {author} {\bibfnamefont {J{\"u}rgen}\
  \bibnamefont {Lisenfeld}}, \bibinfo {author} {\bibfnamefont {Alexander}\
  \bibnamefont {Bilmes}}, \bibinfo {author} {\bibfnamefont {Anthony}\
  \bibnamefont {Megrant}}, \bibinfo {author} {\bibfnamefont {Rami}\
  \bibnamefont {Barends}}, \bibinfo {author} {\bibfnamefont {Julian}\
  \bibnamefont {Kelly}}, \bibinfo {author} {\bibfnamefont {Paul}\ \bibnamefont
  {Klimov}}, \bibinfo {author} {\bibfnamefont {Georg}\ \bibnamefont {Weiss}},
  \bibinfo {author} {\bibfnamefont {John~M.}\ \bibnamefont {Martinis}}, \ and\
  \bibinfo {author} {\bibfnamefont {Alexey~V.}\ \bibnamefont {Ustinov}},\
  }\bibfield  {title} {\enquote {\bibinfo {title} {Electric field spectroscopy
  of material defects in transmon qubits},}\ }\href {\doibase
  10.1038/s41534-019-0224-1} {\bibfield  {journal} {\bibinfo  {journal} {npj
  Quantum Information}\ }\textbf {\bibinfo {volume} {5}},\ \bibinfo {pages}
  {105} (\bibinfo {year} {2019})}\BibitemShut {NoStop}%
\bibitem [{\citenamefont {Bilmes}\ \emph {et~al.}(2020)\citenamefont {Bilmes},
  \citenamefont {Megrant}, \citenamefont {Klimov}, \citenamefont {Weiss},
  \citenamefont {Martinis}, \citenamefont {Ustinov},\ and\ \citenamefont
  {Lisenfeld}}]{Bilmes_SR_2020}%
  \BibitemOpen
  \bibfield  {author} {\bibinfo {author} {\bibfnamefont {Alexander}\
  \bibnamefont {Bilmes}}, \bibinfo {author} {\bibfnamefont {Anthony}\
  \bibnamefont {Megrant}}, \bibinfo {author} {\bibfnamefont {Paul}\
  \bibnamefont {Klimov}}, \bibinfo {author} {\bibfnamefont {Georg}\
  \bibnamefont {Weiss}}, \bibinfo {author} {\bibfnamefont {John~M.}\
  \bibnamefont {Martinis}}, \bibinfo {author} {\bibfnamefont {Alexey~V.}\
  \bibnamefont {Ustinov}}, \ and\ \bibinfo {author} {\bibfnamefont
  {J{\"u}rgen}\ \bibnamefont {Lisenfeld}},\ }\bibfield  {title} {\enquote
  {\bibinfo {title} {Resolving the positions of defects in superconducting
  quantum bits},}\ }\href {\doibase 10.1038/s41598-020-59749-y} {\bibfield
  {journal} {\bibinfo  {journal} {Scientific Reports}\ }\textbf {\bibinfo
  {volume} {10}},\ \bibinfo {pages} {3090} (\bibinfo {year}
  {2020})}\BibitemShut {NoStop}%
\bibitem [{\citenamefont {Bilmes}(2019)}]{Bilmes_Thesis}%
  \BibitemOpen
  \bibfield  {author} {\bibinfo {author} {\bibfnamefont {Alexander}\
  \bibnamefont {Bilmes}},\ }\emph {\bibinfo {title} {Resolving locations of
  defects in superconducting transmon qubits}},\ \href {\doibase
  10.5445/KSP/1000097557} {Ph.D. thesis},\ \bibinfo  {school} {Karlsruher
  Institut für Technologie (KIT)} (\bibinfo {year} {2019})\BibitemShut
  {NoStop}%
\bibitem [{\citenamefont {Bilmes}\ \emph
  {et~al.}(2021{\natexlab{c}})\citenamefont {Bilmes}, \citenamefont
  {Volosheniuk}, \citenamefont {Ustinov},\ and\ \citenamefont
  {Lisenfeld}}]{Bilmes2021probing}%
  \BibitemOpen
  \bibfield  {author} {\bibinfo {author} {\bibfnamefont {Alexander}\
  \bibnamefont {Bilmes}}, \bibinfo {author} {\bibfnamefont {Serhii}\
  \bibnamefont {Volosheniuk}}, \bibinfo {author} {\bibfnamefont {Alexey~V.}\
  \bibnamefont {Ustinov}}, \ and\ \bibinfo {author} {\bibfnamefont {Jürgen}\
  \bibnamefont {Lisenfeld}},\ }\href@noop {} {\enquote {\bibinfo {title}
  {Probing defect densities at the edges and inside {J}osephson junctions of
  superconducting qubits},}\ } (\bibinfo {year} {2021}{\natexlab{c}}),\ \Eprint
  {http://arxiv.org/abs/2108.06555} {arXiv:2108.06555 [quant-ph]} \BibitemShut
  {NoStop}%
\bibitem [{\citenamefont {Klimov}\ and\ \citenamefont
  {Sanchez-Soto}(2000)}]{Klimov_Sanchez}%
  \BibitemOpen
  \bibfield  {author} {\bibinfo {author} {\bibfnamefont {A.~B.}\ \bibnamefont
  {Klimov}}\ and\ \bibinfo {author} {\bibfnamefont {L.~L.}\ \bibnamefont
  {Sanchez-Soto}},\ }\bibfield  {title} {\enquote {\bibinfo {title} {Method of
  small rotations and effective hamiltonians in nonlinear quantum optics},}\
  }\href {\doibase 10.1103/PhysRevA.61.063802} {\bibfield  {journal} {\bibinfo
  {journal} {Phys. Rev. A}\ }\textbf {\bibinfo {volume} {61}},\ \bibinfo
  {pages} {063802} (\bibinfo {year} {2000})}\BibitemShut {NoStop}%
\bibitem [{\citenamefont {Klimov}\ \emph {et~al.}(2002)\citenamefont {Klimov},
  \citenamefont {Sánchez-Soto}, \citenamefont {Navarro},\ and\ \citenamefont
  {Yustas}}]{Klimov_JMO}%
  \BibitemOpen
  \bibfield  {author} {\bibinfo {author} {\bibfnamefont {A.~B.}\ \bibnamefont
  {Klimov}}, \bibinfo {author} {\bibfnamefont {L.~L.}\ \bibnamefont
  {Sánchez-Soto}}, \bibinfo {author} {\bibfnamefont {A.}~\bibnamefont
  {Navarro}}, \ and\ \bibinfo {author} {\bibfnamefont {E.~C.}\ \bibnamefont
  {Yustas}},\ }\bibfield  {title} {\enquote {\bibinfo {title} {Effective
  {H}amiltonians in quantum optics: a systematic approach},}\ }\href {\doibase
  10.1080/09500340210134675} {\bibfield  {journal} {\bibinfo  {journal}
  {Journal of Modern Optics}\ }\textbf {\bibinfo {volume} {49}},\ \bibinfo
  {pages} {2211--2226} (\bibinfo {year} {2002})}\BibitemShut {NoStop}%
\bibitem [{\citenamefont {Eberly}\ and\ \citenamefont
  {W{\'{o}}dkiewicz}(1977)}]{Eberly1977}%
  \BibitemOpen
  \bibfield  {author} {\bibinfo {author} {\bibfnamefont {J~H}\ \bibnamefont
  {Eberly}}\ and\ \bibinfo {author} {\bibfnamefont {K}~\bibnamefont
  {W{\'{o}}dkiewicz}},\ }\bibfield  {title} {\enquote {\bibinfo {title} {{The
  time-dependent physical spectrum of light*}},}\ }\href
  {https://doi.org/10.1364/JOSA.67.001252} {\bibfield  {journal} {\bibinfo
  {journal} {J. Opt. Soc. Am.}\ }\textbf {\bibinfo {volume} {67}},\ \bibinfo
  {pages} {1252--1261} (\bibinfo {year} {1977})}\BibitemShut {NoStop}%
\bibitem [{\citenamefont {Cresser}(1983)}]{CRESSER_1983}%
  \BibitemOpen
  \bibfield  {author} {\bibinfo {author} {\bibfnamefont {J.D.}\ \bibnamefont
  {Cresser}},\ }\bibfield  {title} {\enquote {\bibinfo {title} {Theory of the
  spectrum of the quantised light field},}\ }\href {\doibase
  https://doi.org/10.1016/0370-1573(83)90120-5} {\bibfield  {journal} {\bibinfo
   {journal} {Physics Reports}\ }\textbf {\bibinfo {volume} {94}},\ \bibinfo
  {pages} {47--110} (\bibinfo {year} {1983})}\BibitemShut {NoStop}%
\bibitem [{\citenamefont {Salado-Mej{\'{\i}}a}\ \emph
  {et~al.}(2021)\citenamefont {Salado-Mej{\'{\i}}a}, \citenamefont
  {Rom{\'{a}}n-Ancheyta}, \citenamefont {Soto-Eguibar},\ and\ \citenamefont
  {Moya-Cessa}}]{Salado_Mej_a_2021}%
  \BibitemOpen
  \bibfield  {author} {\bibinfo {author} {\bibfnamefont {M}~\bibnamefont
  {Salado-Mej{\'{\i}}a}}, \bibinfo {author} {\bibfnamefont {R}~\bibnamefont
  {Rom{\'{a}}n-Ancheyta}}, \bibinfo {author} {\bibfnamefont {F}~\bibnamefont
  {Soto-Eguibar}}, \ and\ \bibinfo {author} {\bibfnamefont {H~M}\ \bibnamefont
  {Moya-Cessa}},\ }\bibfield  {title} {\enquote {\bibinfo {title} {Spectroscopy
  and critical quantum thermometry in the ultrastrong coupling regime},}\
  }\href {\doibase 10.1088/2058-9565/abdca5} {\bibfield  {journal} {\bibinfo
  {journal} {Quantum Science and Technology}\ }\textbf {\bibinfo {volume}
  {6}},\ \bibinfo {pages} {025010} (\bibinfo {year} {2021})}\BibitemShut
  {NoStop}%
\bibitem [{\citenamefont {Rom\'an-Ancheyta}\ \emph {et~al.}(2018)\citenamefont
  {Rom\'an-Ancheyta}, \citenamefont {de~los Santos-S\'anchez}, \citenamefont
  {Horvath},\ and\ \citenamefont {Castro-Beltr\'an}}]{RicTime}%
  \BibitemOpen
  \bibfield  {author} {\bibinfo {author} {\bibfnamefont {R.}~\bibnamefont
  {Rom\'an-Ancheyta}}, \bibinfo {author} {\bibfnamefont {O.}~\bibnamefont
  {de~los Santos-S\'anchez}}, \bibinfo {author} {\bibfnamefont
  {L.}~\bibnamefont {Horvath}}, \ and\ \bibinfo {author} {\bibfnamefont
  {H.~M.}\ \bibnamefont {Castro-Beltr\'an}},\ }\bibfield  {title} {\enquote
  {\bibinfo {title} {Time-dependent spectra of a three-level atom in the
  presence of electron shelving},}\ }\href {\doibase
  10.1103/PhysRevA.98.013820} {\bibfield  {journal} {\bibinfo  {journal} {Phys.
  Rev. A}\ }\textbf {\bibinfo {volume} {98}},\ \bibinfo {pages} {013820}
  (\bibinfo {year} {2018})}\BibitemShut {NoStop}%
\bibitem [{\citenamefont {Rom{\'{a}}n-Ancheyta}\ \emph
  {et~al.}(2020)\citenamefont {Rom{\'{a}}n-Ancheyta}, \citenamefont
  {{\c{C}}akmak},\ and\ \citenamefont
  {Müstecapl{\i}o{\u{g}}lu}}]{roman2019spectral}%
  \BibitemOpen
  \bibfield  {author} {\bibinfo {author} {\bibfnamefont {Ricardo}\ \bibnamefont
  {Rom{\'{a}}n-Ancheyta}}, \bibinfo {author} {\bibfnamefont {Bar{\i}{\c{s}}}\
  \bibnamefont {{\c{C}}akmak}}, \ and\ \bibinfo {author} {\bibfnamefont
  {\"Ozg\"ur~E}\ \bibnamefont {Müstecapl{\i}o{\u{g}}lu}},\ }\bibfield  {title}
  {\enquote {\bibinfo {title} {Spectral signatures of non-thermal baths in
  quantum thermalization},}\ }\href {\doibase 10.1088/2058-9565/ab5e4f}
  {\bibfield  {journal} {\bibinfo  {journal} {Quantum Science and Technology}\
  }\textbf {\bibinfo {volume} {5}},\ \bibinfo {pages} {015003} (\bibinfo {year}
  {2020})}\BibitemShut {NoStop}%
\bibitem [{\citenamefont {Phillips}(1972)}]{Phillips1972}%
  \BibitemOpen
  \bibfield  {author} {\bibinfo {author} {\bibfnamefont {{W}.~A.}\ \bibnamefont
  {Phillips}},\ }\bibfield  {title} {\enquote {\bibinfo {title} {Tunneling
  states in amorphous solids},}\ }\href {\doibase 10.1007/BF00660072}
  {\bibfield  {journal} {\bibinfo  {journal} {Journal of Low Temperature
  Physics}\ }\textbf {\bibinfo {volume} {7}},\ \bibinfo {pages} {351--360}
  (\bibinfo {year} {1972})}\BibitemShut {NoStop}%
\bibitem [{\citenamefont {Anderson}\ \emph {et~al.}(1972)\citenamefont
  {Anderson}, \citenamefont {Halperin},\ and\ \citenamefont
  {Varma}}]{Halperin_1972}%
  \BibitemOpen
  \bibfield  {author} {\bibinfo {author} {\bibfnamefont {P.~W.}\ \bibnamefont
  {Anderson}}, \bibinfo {author} {\bibfnamefont {B.~I.}\ \bibnamefont
  {Halperin}}, \ and\ \bibinfo {author} {\bibfnamefont {C.~M.}\ \bibnamefont
  {Varma}},\ }\bibfield  {title} {\enquote {\bibinfo {title} {Anomalous
  low-temperature thermal properties of glasses and spin glasses},}\ }\href
  {\doibase 10.1080/14786437208229210} {\bibfield  {journal} {\bibinfo
  {journal} {The Philosophical Magazine: A Journal of Theoretical Experimental
  and Applied Physics}\ }\textbf {\bibinfo {volume} {25}},\ \bibinfo {pages}
  {1--9} (\bibinfo {year} {1972})}\BibitemShut {NoStop}%
\bibitem [{\citenamefont {Yu}\ and\ \citenamefont
  {Carruzzo}(2021)}]{Carruzzo_2021}%
  \BibitemOpen
  \bibfield  {author} {\bibinfo {author} {\bibfnamefont {Clare~C.}\
  \bibnamefont {Yu}}\ and\ \bibinfo {author} {\bibfnamefont {Herv\'e~M.}\
  \bibnamefont {Carruzzo}},\ }\href@noop {} {\enquote {\bibinfo {title}
  {Two-level systems and the tunneling model: A critical view},}\ } (\bibinfo
  {year} {2021}),\ \Eprint {http://arxiv.org/abs/2101.02787} {arXiv:2101.02787
  [cond-mat.dis-nn]} \BibitemShut {NoStop}%
\bibitem [{\citenamefont {Klein}\ \emph {et~al.}(1978)\citenamefont {Klein},
  \citenamefont {Fischer}, \citenamefont {Anderson},\ and\ \citenamefont
  {Anthony}}]{Klein_PRB_1978}%
  \BibitemOpen
  \bibfield  {author} {\bibinfo {author} {\bibfnamefont {Michael~W.}\
  \bibnamefont {Klein}}, \bibinfo {author} {\bibfnamefont {Baruch}\
  \bibnamefont {Fischer}}, \bibinfo {author} {\bibfnamefont {A.~C.}\
  \bibnamefont {Anderson}}, \ and\ \bibinfo {author} {\bibfnamefont {P.~J.}\
  \bibnamefont {Anthony}},\ }\bibfield  {title} {\enquote {\bibinfo {title}
  {Strain interactions and the low-temperature properties of glasses},}\ }\href
  {\doibase 10.1103/PhysRevB.18.5887} {\bibfield  {journal} {\bibinfo
  {journal} {Phys. Rev. B}\ }\textbf {\bibinfo {volume} {18}},\ \bibinfo
  {pages} {5887--5891} (\bibinfo {year} {1978})}\BibitemShut {NoStop}%
\bibitem [{\citenamefont {{Frisk Kockum}}\ \emph {et~al.}(2019)\citenamefont
  {{Frisk Kockum}}, \citenamefont {Miranowicz}, \citenamefont {{De Liberato}},
  \citenamefont {Savasta},\ and\ \citenamefont {Nori}}]{FriskKockum2019}%
  \BibitemOpen
  \bibfield  {author} {\bibinfo {author} {\bibfnamefont {Anton}\ \bibnamefont
  {{Frisk Kockum}}}, \bibinfo {author} {\bibfnamefont {Adam}\ \bibnamefont
  {Miranowicz}}, \bibinfo {author} {\bibfnamefont {Simone}\ \bibnamefont {{De
  Liberato}}}, \bibinfo {author} {\bibfnamefont {Salvatore}\ \bibnamefont
  {Savasta}}, \ and\ \bibinfo {author} {\bibfnamefont {Franco}\ \bibnamefont
  {Nori}},\ }\bibfield  {title} {\enquote {\bibinfo {title} {{Ultrastrong
  coupling between light and matter}},}\ }\href {\doibase
  10.1038/s42254-018-0006-2} {\bibfield  {journal} {\bibinfo  {journal} {Nature
  Reviews Physics}\ }\textbf {\bibinfo {volume} {1}},\ \bibinfo {pages}
  {19--40} (\bibinfo {year} {2019})},\ \Eprint
  {http://arxiv.org/abs/1807.11636} {1807.11636} \BibitemShut {NoStop}%
\bibitem [{\citenamefont {Scali}\ \emph {et~al.}(2021)\citenamefont {Scali},
  \citenamefont {Anders},\ and\ \citenamefont {Correa}}]{Correa_Quantum_2021}%
  \BibitemOpen
  \bibfield  {author} {\bibinfo {author} {\bibfnamefont {Stefano}\ \bibnamefont
  {Scali}}, \bibinfo {author} {\bibfnamefont {Janet}\ \bibnamefont {Anders}}, \
  and\ \bibinfo {author} {\bibfnamefont {Luis~A.}\ \bibnamefont {Correa}},\
  }\bibfield  {title} {\enquote {\bibinfo {title} {Local master equations
  bypass the secular approximation},}\ }\href {\doibase
  10.22331/q-2021-05-01-451} {\bibfield  {journal} {\bibinfo  {journal}
  {{Quantum}}\ }\textbf {\bibinfo {volume} {5}},\ \bibinfo {pages} {451}
  (\bibinfo {year} {2021})}\BibitemShut {NoStop}%
\bibitem [{\citenamefont {Cattaneo}\ \emph {et~al.}(2019)\citenamefont
  {Cattaneo}, \citenamefont {Giorgi}, \citenamefont {Maniscalco},\ and\
  \citenamefont {Zambrini}}]{Cattaneo_2019}%
  \BibitemOpen
  \bibfield  {author} {\bibinfo {author} {\bibfnamefont {Marco}\ \bibnamefont
  {Cattaneo}}, \bibinfo {author} {\bibfnamefont {Gian~Luca}\ \bibnamefont
  {Giorgi}}, \bibinfo {author} {\bibfnamefont {Sabrina}\ \bibnamefont
  {Maniscalco}}, \ and\ \bibinfo {author} {\bibfnamefont {Roberta}\
  \bibnamefont {Zambrini}},\ }\bibfield  {title} {\enquote {\bibinfo {title}
  {Local versus global master equation with common and separate baths:
  superiority of the global approach in partial secular approximation},}\
  }\href {\doibase 10.1088/1367-2630/ab54ac} {\bibfield  {journal} {\bibinfo
  {journal} {New Journal of Physics}\ }\textbf {\bibinfo {volume} {21}},\
  \bibinfo {pages} {113045} (\bibinfo {year} {2019})}\BibitemShut {NoStop}%
\bibitem [{\citenamefont {Lisenfeld}\ \emph {et~al.}(2016)\citenamefont
  {Lisenfeld}, \citenamefont {Bilmes}, \citenamefont {Matityahu}, \citenamefont
  {Zanker}, \citenamefont {Marthaler}, \citenamefont {Schechter}, \citenamefont
  {Sch{\"o}n}, \citenamefont {Shnirman}, \citenamefont {Weiss},\ and\
  \citenamefont {Ustinov}}]{Lisenfeld2016SR}%
  \BibitemOpen
  \bibfield  {author} {\bibinfo {author} {\bibfnamefont {J{\"u}rgen}\
  \bibnamefont {Lisenfeld}}, \bibinfo {author} {\bibfnamefont {Alexander}\
  \bibnamefont {Bilmes}}, \bibinfo {author} {\bibfnamefont {Shlomi}\
  \bibnamefont {Matityahu}}, \bibinfo {author} {\bibfnamefont {Sebastian}\
  \bibnamefont {Zanker}}, \bibinfo {author} {\bibfnamefont {Michael}\
  \bibnamefont {Marthaler}}, \bibinfo {author} {\bibfnamefont {Moshe}\
  \bibnamefont {Schechter}}, \bibinfo {author} {\bibfnamefont {Gerd}\
  \bibnamefont {Sch{\"o}n}}, \bibinfo {author} {\bibfnamefont {Alexander}\
  \bibnamefont {Shnirman}}, \bibinfo {author} {\bibfnamefont {Georg}\
  \bibnamefont {Weiss}}, \ and\ \bibinfo {author} {\bibfnamefont {Alexey~V.}\
  \bibnamefont {Ustinov}},\ }\bibfield  {title} {\enquote {\bibinfo {title}
  {Decoherence spectroscopy with individual two-level tunneling defects},}\
  }\href {\doibase 10.1038/srep23786} {\bibfield  {journal} {\bibinfo
  {journal} {Scientific Reports}\ }\textbf {\bibinfo {volume} {6}},\ \bibinfo
  {pages} {23786} (\bibinfo {year} {2016})}\BibitemShut {NoStop}%
\bibitem [{\citenamefont {Bilmes}\ \emph {et~al.}(2017)\citenamefont {Bilmes},
  \citenamefont {Zanker}, \citenamefont {Heimes}, \citenamefont {Marthaler},
  \citenamefont {Sch\"on}, \citenamefont {Weiss}, \citenamefont {Ustinov},\
  and\ \citenamefont {Lisenfeld}}]{ElectronicDecoherence}%
  \BibitemOpen
  \bibfield  {author} {\bibinfo {author} {\bibfnamefont {Alexander}\
  \bibnamefont {Bilmes}}, \bibinfo {author} {\bibfnamefont {Sebastian}\
  \bibnamefont {Zanker}}, \bibinfo {author} {\bibfnamefont {Andreas}\
  \bibnamefont {Heimes}}, \bibinfo {author} {\bibfnamefont {Michael}\
  \bibnamefont {Marthaler}}, \bibinfo {author} {\bibfnamefont {Gerd}\
  \bibnamefont {Sch\"on}}, \bibinfo {author} {\bibfnamefont {Georg}\
  \bibnamefont {Weiss}}, \bibinfo {author} {\bibfnamefont {Alexey~V.}\
  \bibnamefont {Ustinov}}, \ and\ \bibinfo {author} {\bibfnamefont {J\"urgen}\
  \bibnamefont {Lisenfeld}},\ }\bibfield  {title} {\enquote {\bibinfo {title}
  {Electronic decoherence of two-level systems in a {J}osephson junction},}\
  }\href {\doibase 10.1103/PhysRevB.96.064504} {\bibfield  {journal} {\bibinfo
  {journal} {Phys. Rev. B}\ }\textbf {\bibinfo {volume} {96}},\ \bibinfo
  {pages} {064504} (\bibinfo {year} {2017})}\BibitemShut {NoStop}%
\bibitem [{\citenamefont {Gonz\'alez}\ \emph {et~al.}(2017)\citenamefont
  {Gonz\'alez}, \citenamefont {Correa}, \citenamefont {Nocerino}, \citenamefont
  {Palao}, \citenamefont {Alonso},\ and\ \citenamefont
  {Adesso}}]{Correa_OSIF_2017}%
  \BibitemOpen
  \bibfield  {author} {\bibinfo {author} {\bibfnamefont {J.~Onam}\ \bibnamefont
  {Gonz\'alez}}, \bibinfo {author} {\bibfnamefont {Luis~A.}\ \bibnamefont
  {Correa}}, \bibinfo {author} {\bibfnamefont {Giorgio}\ \bibnamefont
  {Nocerino}}, \bibinfo {author} {\bibfnamefont {Jos\'e~P.}\ \bibnamefont
  {Palao}}, \bibinfo {author} {\bibfnamefont {Daniel}\ \bibnamefont {Alonso}},
  \ and\ \bibinfo {author} {\bibfnamefont {Gerardo}\ \bibnamefont {Adesso}},\
  }\bibfield  {title} {\enquote {\bibinfo {title} {Testing the validity of the
  ‘local’ and ‘global’ {GKLS} master equations on an exactly solvable
  model},}\ }\href {\doibase 10.1142/S1230161217400108} {\bibfield  {journal}
  {\bibinfo  {journal} {Open Systems \& Information Dynamics}\ }\textbf
  {\bibinfo {volume} {24}},\ \bibinfo {pages} {1740010} (\bibinfo {year}
  {2017})}\BibitemShut {NoStop}%
\bibitem [{\citenamefont {Eberly}\ \emph {et~al.}(1980)\citenamefont {Eberly},
  \citenamefont {Kunasz},\ and\ \citenamefont {W{\'{o}}dkiewicz}}]{Eberly1980}%
  \BibitemOpen
  \bibfield  {author} {\bibinfo {author} {\bibfnamefont {J.H.}\ \bibnamefont
  {Eberly}}, \bibinfo {author} {\bibfnamefont {C.V.}\ \bibnamefont {Kunasz}}, \
  and\ \bibinfo {author} {\bibfnamefont {K.}~\bibnamefont {W{\'{o}}dkiewicz}},\
  }\bibfield  {title} {\enquote {\bibinfo {title} {Time-dependent spectrum of
  resonance fluorescence},}\ }\href
  {https://doi.org/10.1088/0022-3700/13/2/011} {\bibfield  {journal} {\bibinfo
  {journal} {Journal of Physics B-Atomic and Molecular Physics}\ }\textbf
  {\bibinfo {volume} {13}},\ \bibinfo {pages} {217--239} (\bibinfo {year}
  {1980})}\BibitemShut {NoStop}%
\bibitem [{\citenamefont {Black}\ and\ \citenamefont
  {Fulde}(1979)}]{Metallic_Glasses}%
  \BibitemOpen
  \bibfield  {author} {\bibinfo {author} {\bibfnamefont {J.~L.}\ \bibnamefont
  {Black}}\ and\ \bibinfo {author} {\bibfnamefont {P.}~\bibnamefont {Fulde}},\
  }\bibfield  {title} {\enquote {\bibinfo {title} {Influence of the
  superconducting state upon the low-temperature properties of metallic
  glasses},}\ }\href {\doibase 10.1103/PhysRevLett.43.453} {\bibfield
  {journal} {\bibinfo  {journal} {Phys. Rev. Lett.}\ }\textbf {\bibinfo
  {volume} {43}},\ \bibinfo {pages} {453--456} (\bibinfo {year}
  {1979})}\BibitemShut {NoStop}%
\bibitem [{\citenamefont {Matityahu}\ \emph {et~al.}(2016)\citenamefont
  {Matityahu}, \citenamefont {Shnirman}, \citenamefont {Sch\"on},\ and\
  \citenamefont {Schechter}}]{PRB_spectral_diffusion_16}%
  \BibitemOpen
  \bibfield  {author} {\bibinfo {author} {\bibfnamefont {Shlomi}\ \bibnamefont
  {Matityahu}}, \bibinfo {author} {\bibfnamefont {Alexander}\ \bibnamefont
  {Shnirman}}, \bibinfo {author} {\bibfnamefont {Gerd}\ \bibnamefont
  {Sch\"on}}, \ and\ \bibinfo {author} {\bibfnamefont {Moshe}\ \bibnamefont
  {Schechter}},\ }\bibfield  {title} {\enquote {\bibinfo {title} {Decoherence
  of a quantum two-level system by spectral diffusion},}\ }\href {\doibase
  10.1103/PhysRevB.93.134208} {\bibfield  {journal} {\bibinfo  {journal} {Phys.
  Rev. B}\ }\textbf {\bibinfo {volume} {93}},\ \bibinfo {pages} {134208}
  (\bibinfo {year} {2016})}\BibitemShut {NoStop}%
\bibitem [{\citenamefont {Mei\ss{}ner}\ \emph {et~al.}(2018)\citenamefont
  {Mei\ss{}ner}, \citenamefont {Seiler}, \citenamefont {Lisenfeld},
  \citenamefont {Ustinov},\ and\ \citenamefont
  {Weiss}}]{tunneling_fluctuators}%
  \BibitemOpen
  \bibfield  {author} {\bibinfo {author} {\bibfnamefont {Saskia~M.}\
  \bibnamefont {Mei\ss{}ner}}, \bibinfo {author} {\bibfnamefont {Arnold}\
  \bibnamefont {Seiler}}, \bibinfo {author} {\bibfnamefont {J\"urgen}\
  \bibnamefont {Lisenfeld}}, \bibinfo {author} {\bibfnamefont {Alexey~V.}\
  \bibnamefont {Ustinov}}, \ and\ \bibinfo {author} {\bibfnamefont {Georg}\
  \bibnamefont {Weiss}},\ }\bibfield  {title} {\enquote {\bibinfo {title}
  {Probing individual tunneling fluctuators with coherently controlled
  tunneling systems},}\ }\href {\doibase 10.1103/PhysRevB.97.180505} {\bibfield
   {journal} {\bibinfo  {journal} {Phys. Rev. B}\ }\textbf {\bibinfo {volume}
  {97}},\ \bibinfo {pages} {180505} (\bibinfo {year} {2018})}\BibitemShut
  {NoStop}%
\end{thebibliography}
\end{document}